\documentclass[aps,prd,preprint, onecolumn, tightenlines, notitlepage, superscriptaddress, nofootinbib, preprintnumbers, floatfix]{revtex4-2}

\usepackage{amstext}
\usepackage{amssymb}
\usepackage{amsmath}
\usepackage{graphicx}
\graphicspath{{plots/}}
\usepackage{hyperref}
\usepackage{url}
\usepackage{color}
\usepackage{ulem}
\usepackage[utf8]{inputenc}
\pdfoutput=1
\usepackage[x11names]{xcolor}
\usepackage{textcomp}

\usepackage{epsfig,amsfonts,mathrsfs,amsmath,amssymb,graphicx,color,slashed,multirow}
\usepackage{amsmath,latexsym,amssymb,graphicx,slashed,hyperref,color,enumerate,url,cancel,gensymb}
\usepackage{textcomp}

\hypersetup{colorlinks,citecolor= nicegreen,linkcolor= nicegreen}
\definecolor{nicered}{rgb}{0.7,0.1,0.1}
\definecolor{nicegreen}{rgb}{0.1,0.5,0.1}

\def\beq{\begin{equation}}
\def\eeq{\end{equation}}
\def\beqn{\begin{eqnarray}}
\def\eeqn{\end{eqnarray}}

\definecolor{darkblue}{rgb}{0,0,80}

\AtBeginDocument{\hypersetup{citecolor=Green3,linkcolor=Blue1,urlcolor=Blue1}}

\begin{document}

\title{{\Large Five texture zeros in the lepton sector and neutrino oscillations at DUNE}}

\author{Richard H.~Benavides}\email{richardbenavides@itm.edu.co}
\affiliation{Instituto Tecnol\'{o}gico Metropolitano, Facultad de Ciencias Exactas y Aplicadas, Medell\'{i}n, Colombia.}
\author{D.~V.~Forero}\email{dvanegas@udemedellin.edu.co}
\affiliation{Universidad de Medell\'{i}n, Facultad de Ciencias Básicas, Carrera 87 N° 30 - 65, Medell\'{i}n, Colombia}

\author{Luis Mu\~{n}oz}\email{luismunoz@itm.edu.co}
\affiliation{Instituto Tecnol\'{o}gico Metropolitano, Facultad de Ciencias Exactas y Aplicadas, Medell\'{i}n, Colombia.}

\author{Jose M.~Munoz}\email{jose.munoz25@eia.edu.co}
\affiliation{Universidad EIA, grupo FTA, Envigado, Colombia.}

\author{Alejandro Rico}\email{alejandromejia6237@correo.itm.edu.co}
\affiliation{Instituto Tecnol\'{o}gico Metropolitano, Facultad de Ciencias Exactas y Aplicadas, Medell\'{i}n, Colombia.}

\author{A.~Tapia}\email{atapia@udemedellin.edu.co}
\affiliation{Universidad de Medell\'{i}n, Facultad de Ciencias Básicas, Carrera 87 N° 30 - 65, Medell\'{i}n, Colombia}

\begin{abstract}
In this work, we have assumed special structures for the charged and neutral mass matrices in the lepton sector, inspired by structures for the up and down quark mass matrices that result by assuming a certain number of symmetrical zeros in their entries named {\it texture zeros}. A prediction of the lepton mixing matrix results from the rotation matrices that diagonalize the mass matrices in the neutral and charged lepton sectors. The use of texture zeros reduces the number of spurious parameters to the minimal ones needed to explain observations i.e. charged lepton masses and neutrino oscillation parameters. Specifically, we have considered the case of five texture zeros and we have confronted the resulting lepton mixing matrices with current measurements in the neutrino sector. Finally, sensitivities to the independent parameters in the mixing predicted by the nonequivalent forms were studied using simulated events at the DUNE neutrino oscillation experiment. We have found that DUNE is sensitive to nonzero $CP$ violation allowed in the models.
\end{abstract}

\maketitle
\newpage
\tableofcontents

\section{\label{sec:Intro} Introduction}

The Standard Model (SM) of the strong and electroweak interactions has been successful explaining most of the high-energy physics observations. However, several unanswered questions remain. For instance, there is neither an explanation of dark matter origin nor of the baryon asymmetry of the Universe within the SM~\cite{1937ApJ....86..217Z,doi:10.1146/annurev.aa.25.090187.002233,RevModPhys.90.045002,Barr:1991qn}. Experimentally, neutrino oscillations provide the first evidence the SM is incomplete since neutrinos were assumed to be massless in the first version of the electroweak theory. 

Neutrino oscillations is the leading mechanism behind the neutrino flavor conversion phenomena observed in several experiments for more than two decades~\cite{Super-Kamiokande:1998kpq,SNO:2002tuh,KamLAND:2002uet}. More importantly, neutrino oscillations imply neutrinos are massive and mixed particles. Currently, most of the neutrino oscillation parameters are known with a precision below $\sim 6\%$~\cite{deSalas:2020pgw,Esteban:2020cvm,Capozzi:2021fjo}, except for the unknown phase that encodes the possibility of the violation of the charged parity ($CP$) symmetry in the lepton sector~\cite{Nunokawa:2007qh,Branco:2011zb}, the so called Dirac $CP$ phase. With the current precision, it is not known whether the atmospheric mixing angle is maximal or what its correct octant is. The neutrino mass ordering has not yet been established. Future neutrino oscillation experiments, like DUNE~\cite{DUNE:2020ypp,DUNE:2020jqi}, Hyper-Kamiokande~\cite{Hyper-Kamiokande:2018ofw}, and JUNO~\cite{JUNO:2015zny} are expected to make definitive measurements of the mentioned unknowns and also to improve the precision of the neutrino oscillation parameters to $\sim 1\%$ level. 

Although the mechanism behind neutrino masses is unknown, several realizations of the dimension-five operator~\cite{Weinberg:1980bf} are possible. One of the popular extensions of the SM to generate neutrino masses in the theory is the addition of three right-handed neutrino fields that are singlet under the SM gauge group. In this case, neutrino masses are generated via the seesaw Type-I mechanism~\cite{Gell-Mann:1979vob,Yanagida:1979as,Mohapatra:1979ia,Schechter:1980gr}. Other alternatives are the Type-II~\cite{Schechter:1980gr,Lazarides:1980nt,Mohapatra:1980yp} and Type-III~\cite{Foot:1988aq}, which add $SU(2)_L$ triplet scalars or $SU(2)_L$ triplet fermion fields to the SM, respectively. Another possibility are the radiative neutrino mass models~\cite{Cai:2017jrq,Babu:2019mfe}.

Assuming only Dirac neutrinos, the masses of the neutral leptons can be obtained in the same way as the ones for the charged leptons in the SM~\cite{Donoghue:1992dd}. In this case, texture zeros~\cite{Fritzsch:1977vd,Ramond:1993kv,Leontaris:1995vb,Lola:1998xp,Fritzsch:1999ee,Ponce:2011qp, Ponce:2013nsa,Benavides:2020pjx}, applied to the quark sector, might also be implemented to the lepton sector. Several papers have been published on mass matrices for Dirac neutrinos, see for instance~\cite{Ahuja:2009jj,Verma:2013cza,Singh:2018lao,Benavides:2020pjx,Xing:2020ijf}. It is possible to use a weak basis transformation (WBT) to eliminate some entries of the mass matrices or to find equivalent models with texture zeros~\cite{Fritzsch:1999ee,Branco:1999nb,Ludl:2014axa,Ponce:2011qp,Giraldo:2011ya,Ponce:2013nsa}. 

In the models analyzed here, three right-handed neutrinos are added to the SM, maintaining its same gauge structure, i.e., the right fields are singlets under $SU(2)_L$ symmetry. This condition makes possible the use of the polar decomposition theorem~\cite{prasolov1994problems}, by which it is always possible to decompose a complex matrix as a product of a Hermitian and a unitary matrix; the latter matrix can be absorbed by the redefinition of the fields. This implies that the relevant matrices in the analysis are Hermitian, and therefore, the total number of free parameters is 9 for each mass matrix, i.e., half of the number of free parameters given that a complex $3\times 3$ matrix has 18 real parameters. Notice, however, that in the lepton sector there are eight constraints: the three charged lepton masses, the three mixing angles, the two mass squared differences, plus an unknown Dirac $CP$ phase, assuming only Dirac neutrinos. Therefore, there are 18 degrees of freedom, from both the neutral and the charged lepton sectors, while nine constraints including the Dirac $CP$ phase. The introduction of texture zeros in the mass matrices reduce the number of degrees of freedom in the model. It is then possible to find relations between the constraints and the degrees of freedom, allowing us to make predictions. For example, the mixing angles become a function of the masses of the particles, as was shown in the quark sector. For more details, see for instance Refs~\cite{Ponce:2013nsa,Ponce:2011qp,Giraldo:2011ya}.

Texture zeros can be arbitrarily placed in any of the elements of a mass matrix, because their origin and location in the matrix are unknown. The use of discrete symmetries brings flavor structure to the model providing some guidance in this respect~\cite{Morisi:2012fg,King:2013eh,Xing:2020ijf,Chauhan:2022gkz,Almumin:2022rml}. However, this is not the purpose of the present work and we leave this exploration for future studies. This texture zeros idea was proposed for the first time in the quark sector by Fritzsch with six parallel texture zeros~\cite{Fritzsch:1977vd}, and a texture with six nonparallel zeros was also introduced in Ref.~\cite{He:1989eh}. In 1993 Ramond-Robert-Ross (RRR) proposed five possible texture matrices with five zeros in the quark sector~\cite{Ramond:1993kv}. Here, we analyze the mass matrices inspired by the RRR-forms, applied to the lepton sector, and calculate the predicted lepton mixing.~\footnote{It might also be the case that the lepton mixing matrix structure can not be explained from symmetry assumptions~\cite{deGouvea:2012ac}.} Several phenomenological studies can be then performed using current measurements at neutrino oscillation experiments as well as sensitivity studies at future facilities like the DUNE experiment. 

The paper is organized as follows: In Sec.~\ref{sec:formalism}, we present the forms with five texture zeros inspired by the RRR-forms. It is also shown how, by a WBT, some of the forms are equivalent, i.e., they predict the same neutrino mixing. The detailed analysis of the forms is presented in Sec.~\ref{sec:analysis}, and results for some forms are also presented as Appendixes (\ref{sec:app1} and \ref{sec:app2}). In Sec.~\ref{sec:expfit}, we quantify the DUNE sensitivity to the mixing parameters and to the $CP$ violation within the forms. We summarize and conclude in Sec.\ref{sec:summary}.

\section{\label{sec:formalism} Forms with five texture zeros and the WBT}

The model considered here adds three right-handed neutrinos (SMRHN) to the field content of the SM. After the spontaneous breaking of the local gauge symmetry $SU(2)_L\otimes U(1)_Y \rightarrow U(1)_Q$, the Dirac-Lagrangian mass term for the lepton sector is given by 

\begin{equation}
 -\mathcal{L}=\bar{\nu}_LM_n\nu_R+\bar{l}_LM_ll_R + \text{H.c},
\end{equation}

where $M_n$ and $M_l$ are in general  $3\times3$ complex matrices, and $\nu_{L,R}=(\nu_e,\nu_\mu,\nu_\tau)^T_{L,R}$, $l_{L,R}=(e,\mu,\tau)^T_{L,R}$ (the superscript $T$ stands for transpose).

The weak-current Lagrangian in the interaction basis is

\begin{equation}
 \mathcal{L}_{W^-}=-\frac{g}{\sqrt{2}} W_\mu^{-}\bar{l}^\prime_L\gamma^\mu\nu_L^\prime+\text{H.c},
\end{equation}

changed to the physical basis $\nu_L^\prime=U_\nu\nu_L$ and $l_L^\prime=U_ll_L$, the weak-current Lagrangian is written in the form

\begin{equation}
 \mathcal{L}_{W^-}=-\frac{g}{\sqrt{2}} W_\mu^{-}\bar{l}_L U_l^\dagger U_\nu\gamma^\mu\nu_L+\text{H.c}.
\end{equation}

where the product $U_{PMNS}= U_l^\dagger U_\nu$ is the lepton mixing matrix~\cite{Schechter:1980gr}, called Pontecorvo-Maki-Nakagawa-Sakata matrix (PMNS). 

A WBT is a simultaneous unitary transformation of the mass matrices in both the charged and neutral sectors, leaving the physical parameters invariant~\cite{Branco:1999nb},

\begin{equation}
\begin{split}
M_n \rightarrow M_n^R =U M_n U^\dagger\\
M_l \rightarrow M_l^R=U M_l U^\dagger,
\end{split}
\end{equation}

where $U$ is an arbitrary unitary matrix. Both representations $M_{\nu,l}$ and $M_{\nu,l}^R$ are equivalent, because

\begin{equation}\label{eq:Uinvariant}
  U_{PMNS}=U_l^\dagger  U_\nu=U_l^\dagger U U^\dagger U_\nu=U_l^{R\dagger} U_\nu^R=U_{PMNS}^R
\end{equation}

with $U_\nu^R $ and $U_l^R$ unitary matrices that diagonalize the squared mass matrices  $M_n^RM_n^{R\dagger}$ and $M_l^RM_l^{R\dagger}$, respectively. Using this method, it is possible to assign up to three texture zeros (in the same way as in the quark sector, see Ref.~\cite{Branco:1999nb}) and to find equivalent representations. More texture zeros imply relations between the constraints of the lepton sector.
As we mentioned above, from the Yukawa sector, there are eighteen degrees of freedom. Using the polar representation one can write each of the entries of a Hermitian matrix as twelve modulus and six phases. These twelve parameters must necessary satisfy eight constraints, the charged lepton masses plus the oscillation mixing angles and the two mass-squared differences.  

When five texture zeros are introduced in the Hermitian mass matrices of the neutral and charged lepton sector, only seven positive real parameters remain from the total twelve. It is then possible to have one physical prediction since there are eight constraints. In our case this is the absolute neutrino mass. In the case of the phases, after introducing the five texture zeros, only two out of the six are physical. As it will be shown later, in some cases both phases contribute to the Jarlskog invariant~\cite{Jarlskog:1985ht}, which is invariant under phase redefinitions providing a ruler to quantify the amount of $CP$ violation within the forms.

The five texture zeros can be distributed between the charged and neutral lepton sectors in several ways. For instance, two of them can be assigned to the neutral lepton sector, and the remaining three to the charged lepton sector. With this assignment, it is possible to construct twenty different forms for mass matrices with three texture zeros, six of which have zero determinant implying that one of the masses is equal to zero (a case that we do not consider), and fifteen with two texture zeros. Therefore, in total, there are $14\times 15=210$ possible forms with five texture zeros. However, many of them are related through a WBT transformation. It is also possible to assign three texture zeros to the neutral sector and the remaining two to the charged lepton sector thus doubling the number of possible forms. It is worth mentioning that the main purpose of this paper is not to work out all nonequivalent forms~\cite{Ludl:2014axa}\footnote{This was the subject of Ref.~\cite{Zhou:2004wz} where six texture zeros were considered, before the measurement of a nonzero reactor mixing angle.}. In our case, we consider as reference the RRR-forms for the quark sector~\cite{Ramond:1993kv} applied to the lepton sector. In this case, the number of zeros can be switched between the two sectors. Here we consider the case of two zeros in the neutral sector and three in the charged sector. This choice is convenient since the charged sector is well-constrained while there is freedom in neutral sector to have a free parameter to be adjusted. Specifically, the textures considered in this work are the ones shown in Table~\ref{tab:rrr}. Along this work we assumed normal neutrino mass ordering given that the masses in the quark sector are hierarchical.

\begin{widetext}

\begin{table}[t]
\begin{center}
\begin{tabular}{| c | c | c | }
\hline
Form & $M_n$ & $M_l$   \\ [1ex] \hline
 \hline 
 & & \\
 $\text{RRR}_{1}$ &
$\begin{pmatrix} 0 &  |b_n| \,\, e^{i \alpha_1} & 0 \\ |b_n| \,\, e^{-i \alpha_1} & c_n &  |d_n| \,\, e^{i \alpha_2} \\ 0 & |d_n| \,\, e^{-i \alpha_2} &  a_n  \end{pmatrix}$  & 

$\begin{pmatrix} 0 & |b_l| \,\, e^{i \beta_1} & 0 \\  |b_l| \,\, e^{-i \beta_1} & c_l & 0 \\ 0 & 0  & a_l  \end{pmatrix}$\\ &&\\ \hline &&\\

$\text{RRR}_{3}$ & 
$\begin{pmatrix} 0 &  |b_n| \,\, e^{i \alpha_1} & 0 \\ |b_n| \,\, e^{-i \alpha_1} & c_n &  |d_n| \,\, e^{i \alpha_2} \\ 0 & |d_n| \,\, e^{-i \alpha_2} &  a_n  \end{pmatrix}$  & 

$\begin{pmatrix} 0 & |c_l| \,\, e^{i \beta_1} & 0 \\  |c_l| \,\, e^{-i \beta_1} & 0 & |b_l| \,\, e^{i \beta_2} \\ 0 & |b_l| \,\, e^{-i \beta_2}  & a_l  \end{pmatrix}$\\ &&\\ \hline

 &&\\$\text{RRR}_{4}$ &

$\begin{pmatrix} 0 & 0 & |b_n| \,\, e^{i \alpha_1} \\  0 & c_n & |d_n| \,\, e^{i \alpha_2} \\ |b_n| \,\, e^{-i \alpha_1} & |d_n| \,\, e^{-i \alpha_2}  & a_n  \end{pmatrix}$  &

$\begin{pmatrix} 0 & |b_l| \,\, e^{i \beta_1} & 0 \\  |b_l| \,\, e^{-i \beta_1} & c_l & 0 \\ 0 & 0  & a_l  \end{pmatrix}$\\ &&\\  \hline &&\\

$\text{T}_{1}$ & 
$\begin{pmatrix} 0 & 0 & |b_n| \,\, e^{i \alpha_1} \\  0 & c_n & |d_n| \,\, e^{i \alpha_2} \\ |b_n| \,\, e^{-i \alpha_1} & |d_n| \,\, e^{-i \alpha_2}  & a_n  \end{pmatrix}$  & 

$\begin{pmatrix} 0 & |c_l| \,\, e^{i \beta_1} & 0 \\  |c_l| \,\, e^{-i \beta_1} & 0 & |b_l| \,\, e^{i \beta_2} \\ 0 & |b_l| \,\, e^{-i \beta_2}  & a_l  \end{pmatrix}$ \\ &&\\ \hline 

\hline

\end{tabular}
\caption{Nonequivalent five texture zeros inspired by the Ramond-Robert-Ross forms~\cite{Ramond:1993kv}. The $\text{T}_1$-form is a different new proposal.}
\label{tab:rrr}
\end{center}
\end{table}

\end{widetext}

It is possible to have an equivalent ansatz, i.e., it is possible to change the place of the texture zeros in the mass matrices using a WBT, and the mixing does not change. For example, a simple WBT can be performed using the matrices shown in Table~\ref{tab:swb}~\cite{Giraldo:2011ya}. Notice that they are not the unique unitary matrices, instead, they are simpler unitary matrices that we use to illustrate the concept. For instance, with the assignment of the zeros followed in this paper, two of the forms in Ref.~\cite{Ramond:1993kv} are equivalent. The $\text{RRR}_2$-form is equivalent to the $\text{RRR}_1$ and the $\text{RRR}_5$ is equivalent to the $\text{RRR}_4$-form. The nonequivalent forms appear in Table~\ref{tab:rrr} including a new proposal, the $\text{T}_1$-form.

\begin{widetext}

\begin{table}[t]
\begin{center}
\begin{tabular}{|c|c|c|c|c|c| }
\hline & & & & & \\
$\begin{pmatrix} 1 & 0 & 0 \\ 0 & 1 & 0 \\0 & 0 & 1 \end{pmatrix}$		&  
$\begin{pmatrix} 1 & 0 & 0 \\ 0 & 0 & 1 \\0 & 1 & 0 \end{pmatrix}$	& 
$\begin{pmatrix} 0 & 0 & 1 \\ 0 & 1 & 0 \\1 & 0 & 0 \end{pmatrix}$		& 
$\begin{pmatrix} 0 & 1 & 0 \\ 1 & 0 & 0 \\0 & 0 & 1 \end{pmatrix}$		& 
$\begin{pmatrix} 0 & 0 & 1 \\ 1 & 0 & 0 \\0 & 1 & 0 \end{pmatrix}$		& 
$\begin{pmatrix} 0 & 1 & 0 \\ 0 & 0 & 1 \\1 & 0 & 0 \end{pmatrix}$		\\
& & & & &  \\ \hline 
\end{tabular}
\caption{A short list of unitary matrices that can be used to perform a simple WBT.} 
\label{tab:swb}
\end{center}
\end{table}

\end{widetext}

For example, in Table~\ref{tab:rrr}, the $\text{T}_{1}$-form is given by:
\[M_n'=\begin{pmatrix} 0 & 0 & |b_n|\,\, e^{i \alpha_1}\\ 0 & c_n & |d_n|\,\, e^{i \alpha_2} \\|b_n|\,\, e^{-i \alpha_1} & |d_n |\,\, e^{-i \alpha_2} & a_n \end{pmatrix},\]  \[M_l'=\begin{pmatrix} 0 & |c_l|\,\, e^{i \beta_1} & 0 \\ |c_l|\,\, e^{-i \beta_1}& 0 & |b_l|\,\, e^{i \beta_2} \\0 & |b_l|\,\, e^{-i \beta_2} & a_l \end{pmatrix}.\] \\

Then, one can use the second unitary matrix $P$ from Table~\ref{tab:swb}:

\[P= \begin{pmatrix} 1 \ & 0\  & 0\  \\ 0\  & 0\  & 1\  \\0\  & 1\  & 0\ \end{pmatrix},\]

 to perform a WBT transformation~\cite{Benavides:2020pjx,Giraldo:2011ya} such that $M_n=P M_n' P^T$ and $M_l=P M_l' P^T$, resulting in the following texture:

\begin{equation*}\label{eq:example}
  M_n=  \left(
\begin{array}{ccc}
 0 & |b_n| e^{i\alpha_1}  & 0 \\
|b_n| e^{-i\alpha_1} & a_n & |d_n| e^{i\alpha_2} \\
 0 & |d_n| e^{-i\alpha_2}& c_n \\
\end{array}\right)
\end{equation*}

\begin{equation*} M_l=\left(
\begin{array}{ccc}
 0 & 0 & |c_l| e^{i\beta_1} \\
 0 & a_l & |b_l| e^{i\beta_2} \\
 |c_l| e^{-i\beta_1} & |b_l| e^{-i\beta_2} & 0 \\
\end{array}
\right).
\end{equation*}

The two forms, $\text{T}_1$ and the previous one, are equivalent because the same $U_{PMNS}$ mixing matrix is obtained, as shown in Eq.~(\ref{eq:Uinvariant}). The same is true if any of the $P$ matrices presented in Table~\ref{tab:swb} are used in the WBT transformation. Notice, however, that there might be many other $P$ transformation matrices not listed in Table~\ref{tab:swb}, which implies that the 210 forms do not necessarily produce different lepton mixing since many of them might be related by a WBT transformation.

At this point, it is worth defining a procedure that leads to a general parametrization of the new lepton mixing matrix. To start, the phases in the mass matrices $M$ can be factorized in the form $\Phi M^{\prime} \Phi^*$ such that the $M^{\prime}$ matrix is real~\cite{Ponce:2011qp,Ponce:2013nsa,Benavides:2020pjx}. The phase matrix is defined as a general diagonal matrix of the form $\Phi=\text{Diag}(1,e^{i\phi_1},e^{i\phi_2})$, where $\phi_1$ and $\phi_2$ are thus functions of the $\alpha$ and $\beta$ phases. With this procedure, the real mass matrices are diagonalized by orthogonal rotation matrices. Finally, the lepton mixing matrix is obtained from the multiplication of rotation matrices and the $\Phi$ matrices. For example, let us consider the neutral mass matrix $M_n$ for the $\text{T}_{1}$-form in Table~\ref{tab:rrr}, which can be written in terms of a real mass matrix $M_n^{\prime}$ and the phase $\Phi_n$ matrix, 
\begin{equation}\label{eq:rephasing}
  M_n= \Phi_n M_n' \Phi_n^\dagger,
\end{equation}
with the phase matrix $\Phi_n= \text{Diag}(e^{i\alpha_2},e^{i\alpha_1},e^{i(\alpha_1+\alpha_2)})$, and the real mass matrix,

\begin{equation}
M_n^{\prime}=
\begin{pmatrix} 0 & 0 & |b_n| \\ 
0 & c_n & |d_n| 
\\ |b_n| & |d_n| & a_n 
\end{pmatrix}.
\end{equation}
The same procedure must be applied to the charged mass matrix $M_l$ for the $\text{T}_{1}$-form in Table~\ref{tab:rrr}. One can show that, in this case, the phase matrix is $\Phi_l=\text{Diag}(1,e^{i\beta_1},1)$. The real rotation matrices, $R_l$ and $R_n$, that diagonalize each sector can be found using three invariants, such as the determinant, the trace, and the trace of the mass matrices squared applied to the diagonalization condition~\cite{Branco:1999nb},
\begin{equation}\label{eq:invariants}
\begin{split}
\text{Det}\{M_{(n,l)}^{\text{diag}}\}&=\text{Det}\{M_{(n,l)}\},\\
\text{Tr}\{M_{(n,l)}^{\text{diag}}\}&=\text{Tr}\{M_{(n,l)}\},\\ \text{Tr}\{[M_{(n,l)}^{\text{diag}}]^2\}&=\text{Tr}\{[M_{(n,l)}]^2\},\
\end{split}    
\end{equation}

where $M_{(n,l)}^{\text{diag}}=\text{Diag}\{m_{(1,e)},-m_{(2,\mu)}, m_{(3, \tau)}\}$, to obtain relations between the parameters in the mass matrices and the masses of the particles~\cite{Benavides:2020pjx}. Finally, the real symmetric mass matrices are diagonalized by real rotation matrices $R$, for each sector, and the lepton mixing matrix can be written as,
\begin{equation}\label{eq:newmixing}
K=R_l \Phi R_n^T,
\end{equation}
where we have defined $\Phi \equiv \Phi_l \Phi_n^\dagger$. For the particular case of the $\text{T}_1$-form, we have,
\begin{equation}\label{eq:phi}
    \Phi=\text{Diag}(1,e^{-i(\alpha_1-\alpha_2+\beta_1)},e^{-i\alpha_1})=\text{Diag}(1,e^{i\phi_1},e^{i\phi_2}),
\end{equation}
where we have factored a global phase. Notice that we have denoted the lepton mixing matrix in Eq.~(\ref{eq:newmixing}) with $K$ to differentiate it from the particle data group (PDG) parametrization although both unitary matrices account for the lepton mixing.

The lepton mixing matrix in Eq.~(\ref{eq:newmixing}) is general, and defines a procedure that can be applied to any of the forms in Table~\ref{tab:rrr}. Along this paper, we will use Eq.~(\ref{eq:newmixing}) and the second quality in Eq.~(\ref{eq:phi}) to calculate the lepton mixing matrix, analytically as well as numerically.

The three mixing angles from the PDG \cite{ParticleDataGroup:2020ssz} parametrization of the lepton mixing matrix are related to the parametrization in Eq.~(\ref{eq:newmixing}), as follows:
\begin{equation}\label{eq:mapping}
  \begin{split}
  \tan \theta_{12}&=|K_{e,2}|/|K_{e,1}|,\\
 \sin \theta_{13}&=|K_{e,3}|,\\
 \tan \theta_{23}&=|K_{\mu,3}|/|K_{\tau,3}|,\\
   \end{split}
\end{equation}
where the elements of the mixing matrix $K$ depends on the model parameters after rotation of the real mass matrices from the forms in Table~\ref{tab:rrr}.\\

For future discussions on the potential violation of the $CP$-symmetry in the lepton sector within the forms, we have used the Jarlskog invariant~\cite{Jarlskog:1985ht},
\begin{equation}\label{eq:Jcp}
J_{CP}=\mathcal{I}\{K_{e1}^*K_{\mu 3}^*K_{e3}\,K_{\mu 1}\},
\end{equation}
which is invariant under phase redefinitions. Equations~(\ref{eq:mapping}) and (\ref{eq:Jcp}) provide a mapping between the PDG parametrization and the parametrization in Eq.~(\ref{eq:newmixing}). This mapping will be useful for constraining the free parameters in the forms with measurements performed at neutrino oscillation experiments, as well as for the sensitivity analyses.
%
\section{\label{sec:analysis} Analysis of the forms}
%

In the numerical analyses we use the values of the charged lepton masses in Table~\ref{tab:Cmasses}. Neutrino masses $m_i$ can be written in terms of the neutrino mass-squared differences and the unknown absolute neutrino mass $m_0$. For normal neutrino mass ordering, we have the following relations:
\begin{equation}\label{eq:numasses}
    \begin{split}
        m_1 &= m_0\,,\\
        m_2 &= \sqrt{m_0^2+\Delta m^2_{21}}\,,\\
        m_3 &= \sqrt{m_0^2+\Delta m^2_{31}}\,,\\
    \end{split}
\end{equation}
where $\Delta m^2_{21}$ ($\Delta m^2_{31}$) is the solar (atmospheric) mass-squared difference. The values of the neutrino oscillation parameters, used in the analyses presented in the following sections, appear in Table~\ref{tab:oscparams}.

\begin{table}[t]
\begin{center}
\begin{tabular}{| c | c | c | }
\hline
$m_{e}$ (MeV) &  $m_{\mu}$ (MeV) & $m_{\tau}$ (MeV)\\ \hline
$0.511$ & $105.658$ & $1776.860$ \\ \hline
\end{tabular}
\caption{\label{tab:Cmasses} Values for the charged lepton masses taken from Ref.~\cite{ParticleDataGroup:2020ssz}.}
\end{center}
\end{table}

\begin{table}[t]
\begin{center}
\begin{tabular}{| c | c | c | c | c | }
\hline
$\sin^2{\theta_{12}}\pm \sigma(\sin^2{\theta_{12}})$ &  $\sin^2{\theta_{13}}\pm\sigma(\sin^2{\theta_{13}})$ & $\sin^2{\theta_{23}}\pm\sigma(\sin^2{\theta_{23}})$ & $\Delta m^2_{21}$ ($\text{eV}^2$) & $\Delta m^2_{31}$ ($\text{eV}^2$) \\ \hline
$0.320\pm 0.016$ & $0.0220\pm 0.0007$ & $0.574\pm 0.014$ & $7.50\times 10^{-5}$ & $2.55\times 10^{-3}$ \\ \hline
\end{tabular}
\caption{\label{tab:oscparams} Central values of the neutrino oscillation parameters, for normal neutrino mass ordering, from Ref.~\cite{deSalas:2020pgw}. The standard deviation has also been included for the mixing angles in the first three columns. }
\end{center}
\end{table}

In the following sections we focus on two forms, one with three degrees of freedom ($\text{RRR}_{4}$-form) and another one with four degrees of freedom ($\text{T}_1$-form). First, the model parameters from these forms are constrained in an experiment independent way using the current knowledge of the neutrino oscillation experiments. In Sec.~\ref{sec:expfit} DUNE sensitivities to the model parameters and to leptonic $CP$ violation are also quantified for these forms. Results for the analyses performed for the remaining forms, $\text{RRR}_{1}$ and $\text{RRR}_{3}$, are presented in the Appendixes~\ref{sec:app1} and \ref{sec:app2}, respectively.

\subsection{Analysis of the $\text{RRR}_{4}$-form }

The $\text{RRR}_{4}$-form was defined in Table~\ref{tab:rrr}. For this particular form, the following relations are obtained after applying Eq.~(\ref{eq:rephasing}) for the mass matrices in each sector, together with the procedure in Eq.~(\ref{eq:invariants}),

\begin{equation}\label{eq:rrr4params}
\begin{split}
c_n &=-a_n + m_1 - m_2 + m_3,\\ |b_n| & =\sqrt{m_1 m_2 m_3/(-a_n + m_1 - m_2 + m_3)},\\ |d_n| &=\sqrt{\frac{(a_n -m_1 + m_2) (a_n - m_1 - m_3) (a_n + m_2 - m_3)}{-a_n + m_1 - m_2 + m_3}},\\
a_l & =m_e - m_\mu , \\
|b_l|&= m_\tau,\\ |c_l| &= \sqrt{m_e m_\mu}\,.
\end{split}
\end{equation}

It is possible to diagonalize the mass matrices using the following orthogonal rotation matrices after replacing the relations of Eq.~(\ref{eq:rrr4params}) in the mass matrix entries of the $\text{RRR}_4$-form, without phases,

\begin{widetext}

\begin{equation}\label{eq:rnRRR4}
R_n=\left(
\begin{array}{ccc}
 \sqrt{\frac{m_2 m_3 (a_n+m_2-m_3)}{(m_1+m_2) (m_1-m_3) (-a_n+m_1-m_2+m_3)}} & -\sqrt{\frac{m_1 (a_n-m_1+m_2) (-a_n+m_1+m_3)}{(m_1+m_2) (m_3-m_1) (-a_n+m_1-m_2+m_3)}} & \sqrt{\frac{m_1 (a_n+m_2-m_3)}{(m_1+m_2) (m_1-m_3)}} \\
 -\sqrt{\frac{m_1 m_3 (-a_n+m_1+m_3)}{(m_1+m_2) (m_2+m_3) (-a_n+m_1-m_2+m_3)}} & -\sqrt{\frac{m_2 (a_n-m_1+m_2) (a_n+m_2-m_3)}{(m_1+m_2) (m_2+m_3) (a_n-m_1+m_2-m_3)}} & \sqrt{\frac{m_2 (-a_n+m_1+m_3)}{(m_1+m_2) (m_2+m_3)}} \\
 \sqrt{\frac{m_1 m_2 (a_n-m_1+m_2)}{(m_3-m_1) (m_2+m_3) (-a_n+m_1-m_2+m_3)}} & \sqrt{\frac{m_3 (a_n-m_1-m_3) (a_n+m_2-m_3)}{(m_1-m_3) (m_2+m_3) (a_n-m_1+m_2-m_3)}} & \sqrt{\frac{m_3 (a_n-m_1+m_2)}{(m_3-m_1) (m_2+m_3)}} \\
\end{array}
\right),
\end{equation}

\end{widetext}

and, 

\begin{equation}\label{eq:rlRRR4}
R_l=\left(
\begin{array}{ccc}
 \sqrt{\frac{m_\mu}{(m_e+m_\mu)}} & \sqrt{\frac{m_e}{(m_e+m_\mu)}} & 0 \\
 -\sqrt{\frac{m_e}{(m_e+m_\mu)}} &\sqrt{\frac{m_\mu}{(m_e+m_\mu)}} & 0 \\
 0 & 0 & 1 \\
\end{array}
\right),
\end{equation}

for the neutral and charged lepton sectors, respectively. To ensure all elements in the $R_n$ matrix are real, the mathematical restriction $m_1-m_2<a_n<m_3-m_2$ must be satisfied. This means that given a constraint on $m_0$, $a_n$ is bounded by the neutrino masses in Eq.~(\ref{eq:numasses}). However, only upper limits from direct mass measurements $m_0<1.1~\text{eV}$ at $90\%$ of confidence level (C.L.) ~\cite{KATRIN:2019yun} and from cosmology $\sum_i m_i <0.12~\text{eV}$ at $95\%$ of C.L.~\cite{Planck:2018vyg} (or $m_0\lesssim 3\times 10^{-2}~\text{eV}$, using the values in Table~\ref{tab:oscparams}), are currently known. In any case, the largest allowed range is obtained when $m_0\ll 1$, and therefore, $-\sqrt{\Delta m^2_{21}}<a_n<\sqrt{\Delta m^2_{31}}-\sqrt{\Delta m^2_{21}}$ or numerically $-8.7\times 10^{-3} < a_n/\text{eV} < 4.2\times 10^{-2}$ using the values in Table~\ref{tab:oscparams}. The mathematical range is not affected by the freedom from the precision on the neutrino mass-squared differences, therefore these parameters are fixed along this work. As it will be shown later, the neutrino mixing angles constrain $a_n$ to a smaller range than the one obtained from the mathematical constraint.

Before proceeding with the first numerical analysis, it is worth defining the degrees of freedom involved in the lepton mixing. In theory, from the rotation matrices in Equations.~(\ref{eq:rnRRR4}) and (\ref{eq:rlRRR4}), and the mixing matrix in Eq.~(\ref{eq:newmixing}), there are four free parameters\footnote{Along this work the charged lepton masses have been fixed to their best-fit values in Table~\ref{tab:Cmasses}, therefore, the freedom present in the forms is the one present in the neutral sector.}, namely $\vec{\lambda}=\{m_0,a_n,\phi_1,\phi_2\}$. However, in the case of the $\text{RRR}_1$ and the $\text{RRR}_4$-forms the lepton mixing matrix is independent of the $\phi_2$ phase. We arrived to this conclusion after performing the analysis and calculating the Jarslkog invariant defined in Eq.~(\ref{eq:Jcp}).

In order to constrain the model parameters $\vec{\lambda}$ in each form, the following statistical test was implemented: 
\begin{equation}\label{eq:chi2NOexp}
    \chi^2(\vec{\lambda})=\sum_{i<j} \left( \frac{\sin^2{\theta_{ij}}-\sin^2{\tilde{\theta}_{ij}}}{\sigma(\sin^2{\theta_{ij}})}  \right)^2, \quad \text{with} \quad i,j=1,2,3,
\end{equation}
where $\sin^2{\tilde{\theta}_{ij}}$ are the mixing angles predicted by the forms in Eq.~(\ref{eq:mapping}), while $\sin^2{\theta_{ij}}$ and $\sigma(\sin^2{\theta_{ij}})$ are the current best-fit values and its one-sigma deviations, respectively (see table~\ref{tab:oscparams}). Notice that the $\chi^2$ function in Eq.~(\ref{eq:chi2NOexp}) is basically the sum of penalties assuming the $\Delta \chi^2$ profiles for each one of the mixing angles is Gaussian. This is not true for the atmospheric mixing angle $\sin^2{\theta_{23}}$, and therefore, we penalize the $\chi^2$ function with the $\Delta \chi^2(\sin^2{\theta_{23}})$ profile reported in Ref.~\cite{deSalas:2020pgw}.

\begin{figure}[t]
    \centering
    \includegraphics[width=\textwidth]{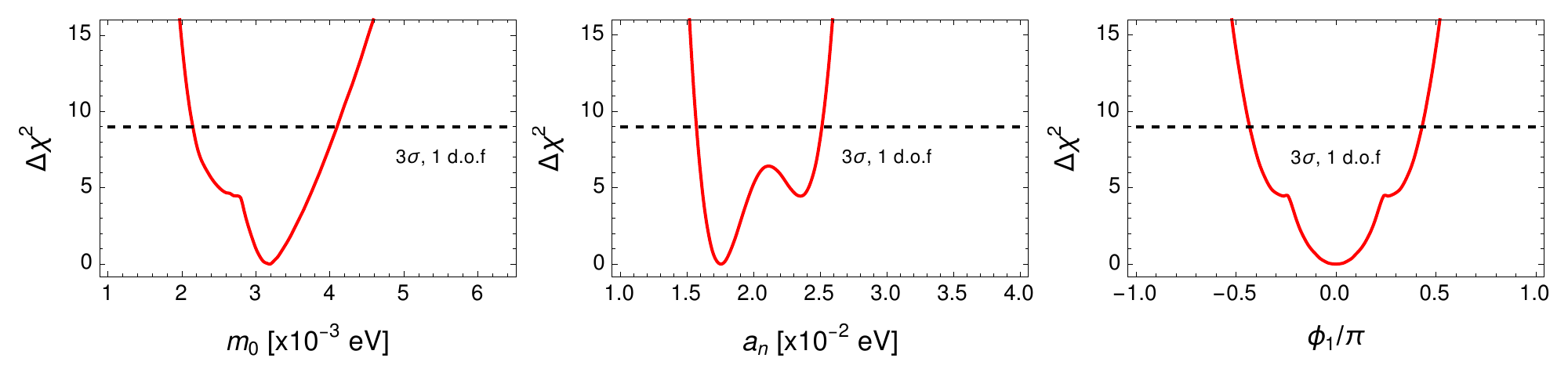}
    \caption{$\Delta \chi^2$ profiles for each one of the three mixing parameters $\vec{\lambda}=\{m_0,a_n,\phi_1\}$ (shown in left, middle and right panels, respectively) that parametrize the lepton mixing obtained from the $\text{RRR}_4$-form, after minimizing over the two parameters not shown in each panel. The horizontal line corresponds to the allowed range at $3\sigma$ of C.L for $1\,\text{d.o.f.}$ whose values are reported in Table~\ref{tab:rrr4fit}. }
    \label{fig:RRR44profiles}
\end{figure}

Figure~\ref{fig:RRR44profiles} shows the $\Delta \chi^2$ profiles for the model parameters, after the minimization of the $\chi^2$ function in Eq.~(\ref{eq:chi2NOexp}). The three parameters are constrained with this analysis. In Table~\ref{tab:rrr4fit} we resume the main results from fit; the best-fit point (bfp) and the three sigma allowed range for each parameter. The value of $\chi^2$ at the bfp $\chi^2_{\text{min.}}$ is given in the caption for completeness.

In this form, the Jarlskog invariant ($J_{\text{CP}}$) only depends on $\phi_1$. We have numerically verified that the $J_{\text{CP}}$ can be written as,
\begin{equation}\label{eq:correspondence}
    J_{\text{CP}}=\pm |J_{\text{CP}}^{\text{max}}|\,\sin{\phi_{1}}\,,
\end{equation}

where the $\phi_1$ phase is the only source of $CP$ violation in the $\text{RRR}_4$-form\footnote{We have numerically verified that for the $\text{RRR}_1$-form the sign in Eq.~(\ref{eq:correspondence}) is plus while for the $\text{RRR}_4$-form is minus.}. From the right panel in Fig.~\ref{fig:RRR44profiles} one can see that values of $\phi_1 \ne 0$ are allowed although the values $\phi_1 =\pm \pi/2$ are excluded at $3\sigma$. From Table~\ref{tab:rrr4fit}, the phase is restricted to the range $[-0.4\pi, \ 0.4\pi]$ at $3\sigma$ of C.L for $1\,\text{d.o.f.}$, implying a restriction on the Jarlskog invariant, $J_{\text{CP}} \in [-9,\ 9]\times 10^{-3}$ at the same C.L. 
Later we will analyze the CP-sensitivity within this form at a future neutrino oscillation experiment given that, currently, a conclusive determination of the Dirac $CP$ phase has not yet been established.

\begin{table}[t]
\begin{center}
\begin{tabular}{| c | c | c | }
\hline
Parameter & Best-fit & $3\sigma$ range \\ \hline \hline
$m_0$ ($\times 10^{-3}\,\text{eV}$) & $3.2$ & $[2.2, 4.1]$ \\ \hline
$a_n$ ($\times 10^{-2}\,\text{eV}$) & $1.8$ & $[1.6, 2.5]$ \\ \hline
$\phi_1/\pi$ & $0$ & $[-0.4, 0.4]$  \\ \hline
$\phi_2/\pi$ & Independent & \\ \hline 
\end{tabular}
\caption{\label{tab:rrr4fit} Best-fit parameters (second column) and three sigma allowed range for $1\,\text{d.o.f.}$ (third column) for each of the parameters of the $\text{RRR}_4$-form (first column) obtained from the minimization of Eq.~(\ref{eq:chi2NOexp}). The $\chi^2$ value at the minimum is $\chi^2_{\text{min.}}=1.4$.} 
\end{center}
\end{table}

\subsection{Analysis of the $\text{T}_{1}$-form }
In contrast to the $\text{RRR}_4$-form analyzed in the previous section, the four parameters $\vec{\lambda}=\{m_0,a_n,\phi_1,\phi_2\}$ contribute to the lepton mixing predicted by this texture. For this particular form, the orthogonal matrix that rotates the neutral mass is the same one obtained for the $\text{RRR}_4$-form, because the same mass matrix structure was assumed in both cases (see table~\ref{tab:rrr}). Thus, the mathematical restriction $m_1-m_2<a_n<m_3-m_2$ must be satisfied implying $-\sqrt{\Delta m^2_{21}}<a_n<\sqrt{\Delta m^2_{31}}-\sqrt{\Delta m^2_{21}}$ or numerically $-8.7\times 10^{-3} < a_n/\text{eV} < 4.2\times 10^{-2}$ using the values in Table~\ref{tab:oscparams}, as discussed in the previous section.

After applying Eq.~(\ref{eq:rephasing}) to the charged mass matrix, together with the procedure in Eq.~(\ref{eq:invariants}), we have found the following relations:

\begin{equation}
\begin{split}
a_l & =m_e - m_\mu + m_\tau, \\
|b_l|&=\sqrt{(m_e - m_\mu) (m_\mu - m_\tau) (m_e + 
   m_\tau)/(m_e - m_\mu + m_\tau)},\\ |c_l| &= \sqrt{(m_e m_\mu m_\tau)/(m_e - m_\mu + m_\tau)}\,,
\end{split}
\end{equation}
   
 and through the same procedure applied to the $\text{RRR}_4$-form, the following orthogonal rotation matrix is obtained:
   
\begin{equation}
R_l=\left(
\begin{array}{ccc}
 -\sqrt{\frac{m_\mu m_\tau (m_\mu-m_\tau)}{(m_e+m_\mu) (m_e-m_\tau) (m_e-m_\mu+m_\tau)}} & -\sqrt{\frac{m_e (m_\mu-m_\tau)}{(m_e+m_\mu) (m_e-m_\tau)}} & \sqrt{\frac{m_e (m_e-m_\mu) (m_e+m_\tau)}{(m_e+m_\mu) (m_e-m_\tau) (m_e-m_\mu+m_\tau)}} \\
 \sqrt{\frac{m_e m_\tau (m_e+m_\tau)}{(m_e+m_\mu) (m_\mu+m_\tau) (m_e-m_\mu+m_\tau)}} & -\sqrt{\frac{m_\mu (m_e+m_\tau)}{(m_e+m_\mu) (m_\mu+m_\tau)}} & \sqrt{\frac{m_\mu (m_e-m_\mu) (m_\mu-m_\tau)}{(m_e+m_\mu) (m_\mu+m_\tau) (m_e-m_\mu+m_\tau)}} \\
 \sqrt{\frac{m_e m_\mu (m_e-m_\mu)}{(m_e-m_\tau) (m_\mu+m_\tau) (m_e-m_\mu+m_\tau)}} & \sqrt{\frac{m_\tau (m_e-m_\mu)}{(m_e-m_\tau) (m_\mu+m_\tau)}} & \sqrt{\frac{m_\tau (m_e+m_\tau) (m_\mu-m_\tau)}{(m_e-m_\tau) (m_\mu+m_\tau) (m_e-m_\mu+m_\tau)}} \\
\end{array}
\right)
\end{equation}

\begin{figure}[t]
    \centering
    \includegraphics[width=\textwidth]{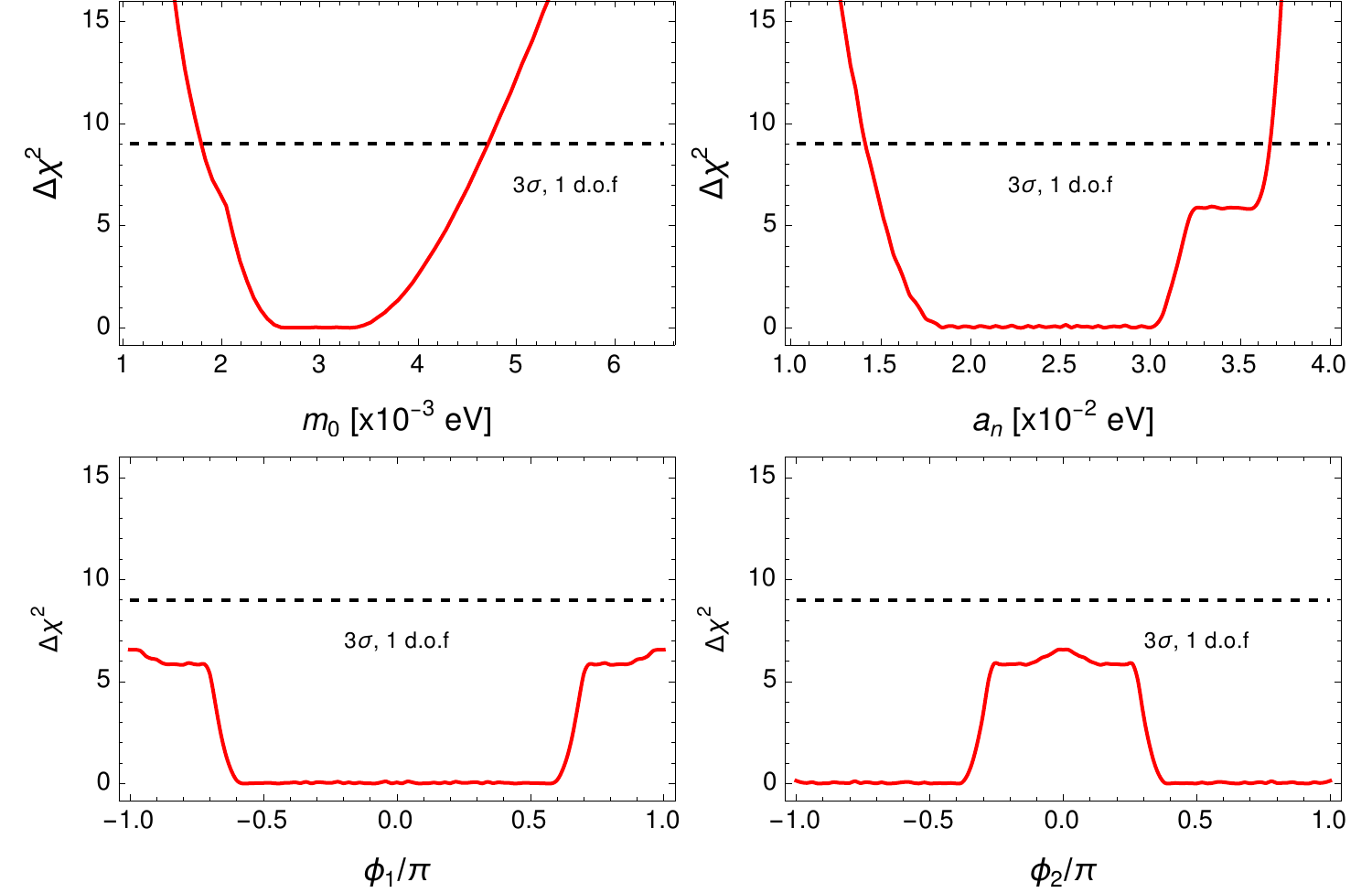}
    \caption{$\Delta \chi^2$ profiles for each one of the four mixing parameters $\vec{\lambda}=\{m_0,a_n,\phi_1, \phi_2\}$ (shown in the upper left, upper right, lower left and lower right panels, respectively) that parametrize the lepton mixing obtained from the $\text{T}_1$-form, after minimizing over the three parameters not shown in each panel. The horizontal line corresponds the allowed range at $3\sigma$ of C.L for $1\,\text{d.o.f.}$ whose values are reported in Table~\ref{tab:t1fit}. }
    \label{fig:T11profiles}
\end{figure}

Figure~\ref{fig:T11profiles} shows the $\Delta \chi^2$ profiles for the model parameters, after the minimization of the $\chi^2$ function in Eq.~(\ref{eq:chi2NOexp}). Two out of the four parameters are constrained with high-confidence, $m_0$ and $a_n$, as seen from the top right and top left panels. In table~\ref{tab:t1fit} we resume the main results from fit: the best-fit point (bfp), the three sigma allowed range for each parameter, and the value of $\chi^2$ at the bfp, $\chi^2_{\text{min.}}$.

\begin{figure}[t]
    \centering
    \includegraphics[width=0.7\textwidth]{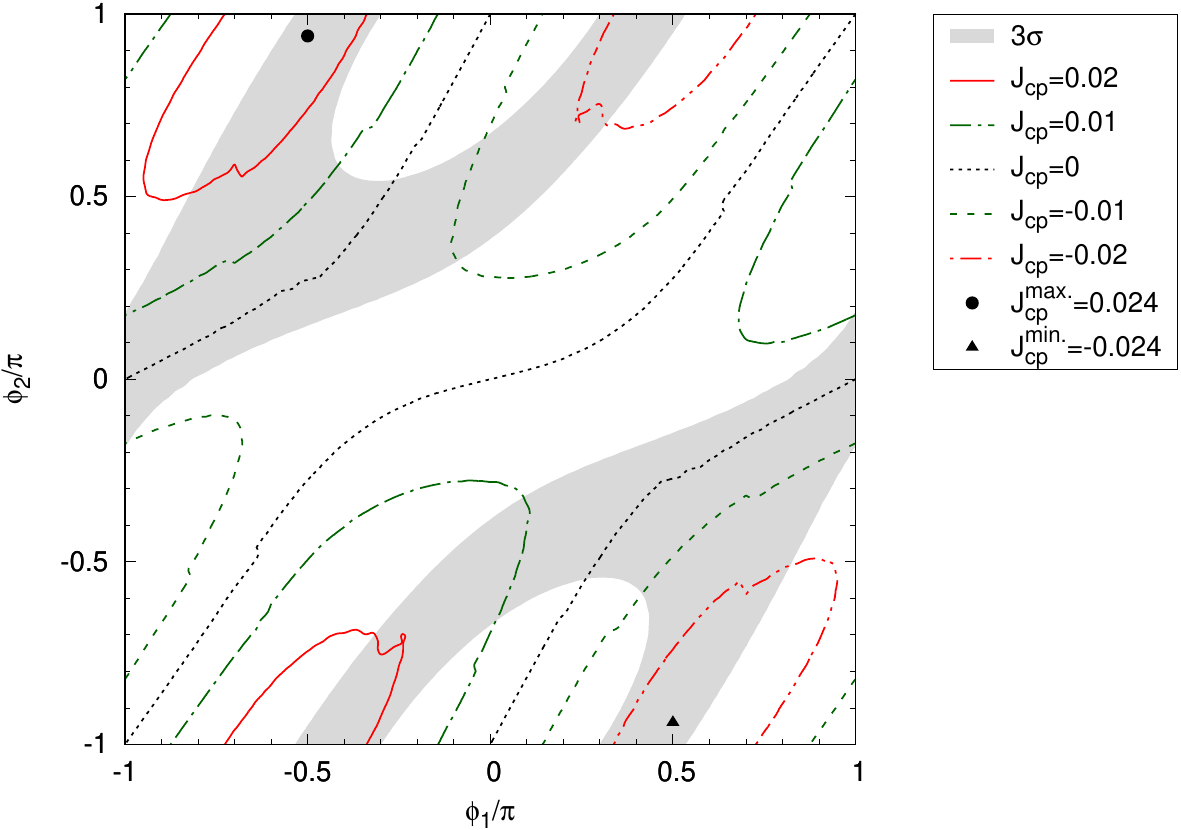}
    \caption{Jarlskog invariant in terms of the $\phi$ phases. The gray region, corresponds to the allowed region at $3\sigma$ of C.L for $1\,\text{d.o.f.}$ after the marginalization of $m_0$ and $a_n$. The Jarlskog is evaluated at these $m_0$ and $a_n$. The lines correspond to some constant values of the Jarlskog. The maximun (minimum) values of the Jarlskog are also labeled with the point (triangle).}
    \label{fig:T11Jcp}
\end{figure}

In this form, the Jarlskog invariant depends on both, $\phi_1$ and $\phi_2$, phases as shown in Fig.~\ref{fig:T11Jcp}. From the bottom right and bottom left panels in Fig.~\ref{fig:T11profiles} one can see the phases are unconstrained at $3\sigma$ of C.L for $1\,\text{d.o.f.}$ and the profiles are flat in most of their parameter space. However, as shown in the gray region of Fig.~\ref{fig:T11Jcp}, both phases are correlated and some part of the $\phi_1-\phi_2$ parameter space is excluded at $3\sigma$ of C.L for $1\,\text{d.o.f.}$. In any case, in the analysis we have only used information on the three mixing angles but fitted four parameters. The situation would change if an additional observable is included, as for instance information on the $CP$-violating phase. In the following section, we analyze all the forms in Table~\ref{tab:rrr} considering a simulation of neutrino events expected at a future neutrino oscillation experiment. We also show how the sensitivity ranges of the physical phases are smaller than the allowed regions for some of the forms.

\begin{table}[t]
\begin{center}
\begin{tabular}{| c | c | c | }
\hline
Parameter & Best-fit & $3\sigma$ range \\ \hline \hline
$m_0$ ($\times 10^{-3}\text{eV}$) & $3.3$ & $[1.8, 4.7]$ \\ \hline
$a_n$ ($\times 10^{-2}\text{eV}$) & $2.3$ & $[1.4, 3.7]$ \\ \hline
$\phi_1/\pi$ & $0.4$ & Unconstrained  \\ \hline
$\phi_2/\pi$ & $0.9$ & Unconstrained  \\ \hline 
\end{tabular}
\caption{\label{tab:t1fit} Best-fit parameters (second column) and three sigma allowed range for $1\,\text{d.o.f.}$ (third column) for each of the parameters of the $T_1$-form (first column) obtained from the minimization of Eq.~(\ref{eq:chi2NOexp}). The $\chi^2$ value at the minimum is $\chi^2_{\text{min.}}=0$.} 
\end{center}
\end{table}

\section{DUNE sensitivity to the mixing parameters}
\label{sec:expfit}

The Deep Underground Neutrino Experiment (DUNE) is a long-baseline multipurpose neutrino oscillation experiment that is under construction. The neutrino beam, produced at Fermilab, is going to be directed towards the far detector (FD) located at the Sanford Underground Research Facility. Hence, the neutrinos produced at Fermilab will travel $\sim 1300$~km before reaching the FD. The neutrino beam is characterized before neutrinos oscillate to reduce beam-related systematical uncertainties by the use of a near detector (ND) complex. Both detectors are time projection chambers filled with liquid argon. The FD will have four modules, each one of 10~kt that will be operational at different stages. For the purpose of this work, we assume a 40~kt FD fiducial mass, a beam power of 1.2~MW, and a running time of 6.5~years in neutrino and antineutrino modes which gives an exposure of 624~MW-kt-yr. We follow the general DUNE set up described in Ref.~\cite{DUNE:2021cuw}.

\subsection{Details of the analysis} 

The lepton mixing matrix, calculated for each form in Table~\ref{tab:rrr} using the general parametrization in Eq.~(\ref{eq:newmixing}), was implemented in the GLoBES C-library~\cite{Huber:2004ka,Huber:2007ji} probability engine.

We have simulated charged current (CC) events at the DUNE FD from the four oscillation channels, electron-(anti)neutrino appearance and muon-(anti)neutrino disappearance, as signal. In the case of the electron-(anti)neutrino appearance oscillation channels, electron-(anti)neutrino disappearance events (from beam contamination), muon-(anti)neutrino disappearance and tau-(anti)neutrino appearance, and neutral current (NC) events make up the simulated background. In the case of the muon-(anti)neutrino disappearance oscillation channels, tau-(anti)neutrino appearance events and neutral current events make up the main background. Detector effects, as efficiencies and energy resolutions, were also included in the calculation of the events using information provided in Ref.~\cite{DUNE:2021cuw}. Muon-(anti)neutrino fluxes at the source, as well as CC and NC cross section interactions in argon at the FD, were also provided in the same reference. A constant matter density profile has been assumed with a density value of $2.9~\text{g/cm}^3$ to account for the neutrino propagation in matter.

To quantify the DUNE sensitivity to the predicted mixing parameters $\vec{\lambda}$, we use the usual $\chi^2$ function for hypothesis testing,
\begin{equation}\label{eq:chi2}
\chi^2(\vec{\theta}_0;\vec{\lambda}; \vec{\xi})=2\sum_{i=1}^{\text{bins}}\left[\mu_i(\vec{\lambda}, \vec{\xi})-n_i(\vec{\theta}_0)+n_i(\vec{\theta}_0)\,\text{ln}\frac{n_i(\vec{\theta}_0)}{\mu_i(\vec{\lambda}, \vec{\xi})}\right]+\chi^2_{\text{syst.}}(\vec{\xi}),
\end{equation}
where $n_i$ are the simulated signal plus background events (simulated data) at each energy bin $i$, assuming the standard parametrization of the lepton mixing matrix (PDG parametrization)~\cite{ParticleDataGroup:2020ssz}, which depends on the input values for the standard neutrino-oscillation parameters $\vec{\theta}_0$. The test signal plus background events $\mu_i$ are calculated assuming the parametrization in Eq.~(\ref{eq:newmixing}) that depends on the $\vec{\lambda}$ and the nuisance parameters $\vec{\xi}$ accounting for the systematical uncertainties. Systematical uncertainties accounting for flux and cross section uncertainties has been included to the analysis in the form of total normalization uncertainties (norm) for each one of the oscillation channels in signal (background) events, ranging from 2\% to 5\% (5\% to 20\%). Besides this systematical uncertainties implemented in the ancillary files released in Ref.~\cite{DUNE:2021cuw}, here we have also included a 5\% bin-to-bin uncorrelated systematical uncertainty (shapelike systematical uncertainties) in the muon-(anti)neutrino disappearance signal in order to properly account for the allowed parameter region for the atmospheric plane (see Fig.~26 in Ref.~\cite{DUNE:2020jqi}). The following penalties are added to Eq.~(\ref{eq:chi2}):
\begin{equation}\label{eq:syst}
\chi^2_{\text{syst.}}(\vec{\xi})=\sum_{k=1}^{\text{channels}}\left(\frac{\xi_{\text{norm}_k}}{\sigma_{\text{norm}_k}}\right)^2+\sum_{l=1}^{\text{bins}}\left(\frac{\xi_{\text{shape}_l}}{\sigma_{\text{shape}_l}}\right)^2\,.
\end{equation}
The DUNE sensitivity to the parameters $\vec{\lambda}$ is obtained from Eq.~(\ref{eq:chi2}) after minimizing over the nuisance parameters,
\begin{equation}\label{eq:totChi2}
\chi^2(\vec{\theta}_0;\vec{\lambda})=\text{min}_{\vec{\xi}}\left.\{\chi^2(\vec{\theta}_0;\vec{\lambda}; \vec{\xi})\right\}\,.
\end{equation}
In the following section we present the results obtained after minimizing the $\chi^2$ function in Eq.~(\ref{eq:totChi2}) over the $\vec{\lambda}=\{m_0,a_n,\phi_1,\phi_2\}$ parameters in terms of input values of the standard oscillation parameters $\vec{\theta}_0=\{\sin^2 \theta_{ij}, \delta_{CP}, \Delta m^2_{k1}\}$. i.e., the three mixing angles $\sin^2 \theta_{ij}$, the Dirac $CP$-violating phase $\delta_{CP}$, and the two mass-squared differences $\Delta m^2_{k1}$ (with $k=2,3$).

\subsection{Results for the $\text{RRR}_4$-form}

In this section we present results for the analysis described in the previous section, testing the mixing resulted from the $\text{RRR}_4$-form in Table~\ref{tab:rrr}, using Eq.~(\ref{eq:newmixing}) and described in more detail in Sec.~\ref{sec:analysis}. The main discussion also applies to any of the forms in Table~\ref{tab:rrr}. However, results for the remaining forms appear in the next section ($\text{T}_1$-form) or have been included in the Appendixes \ref{sec:app1} and \ref{sec:app2} ($\text{RRR}_1$ and $\text{RRR}_3$-forms). 

In all the analyses, we have fixed the standard oscillation parameters to their best-fit point in Table~\ref{tab:oscparams}, except for the atmospheric mixing angle $\theta_{23}$. We consider values of $\theta_{23}$ in the lower ($\theta_{23}<\pi/4$) and upper ($\theta_{23}>\pi/4$) octant as well as maximal mixing ($\theta_{23}=\pi/4$). Input values for the Dirac $CP$ phase $\delta_{\text{CP}}$, consistent with maximal $CP$ violation ($\delta_{\text{CP}}=\pm \pi/2$), as well as $CP$ conserving values ($\delta_{\text{CP}}=\{0,\pm \pi\}$), were also assumed.

\begin{figure}[t]
\centering
    \includegraphics[width=\textwidth]{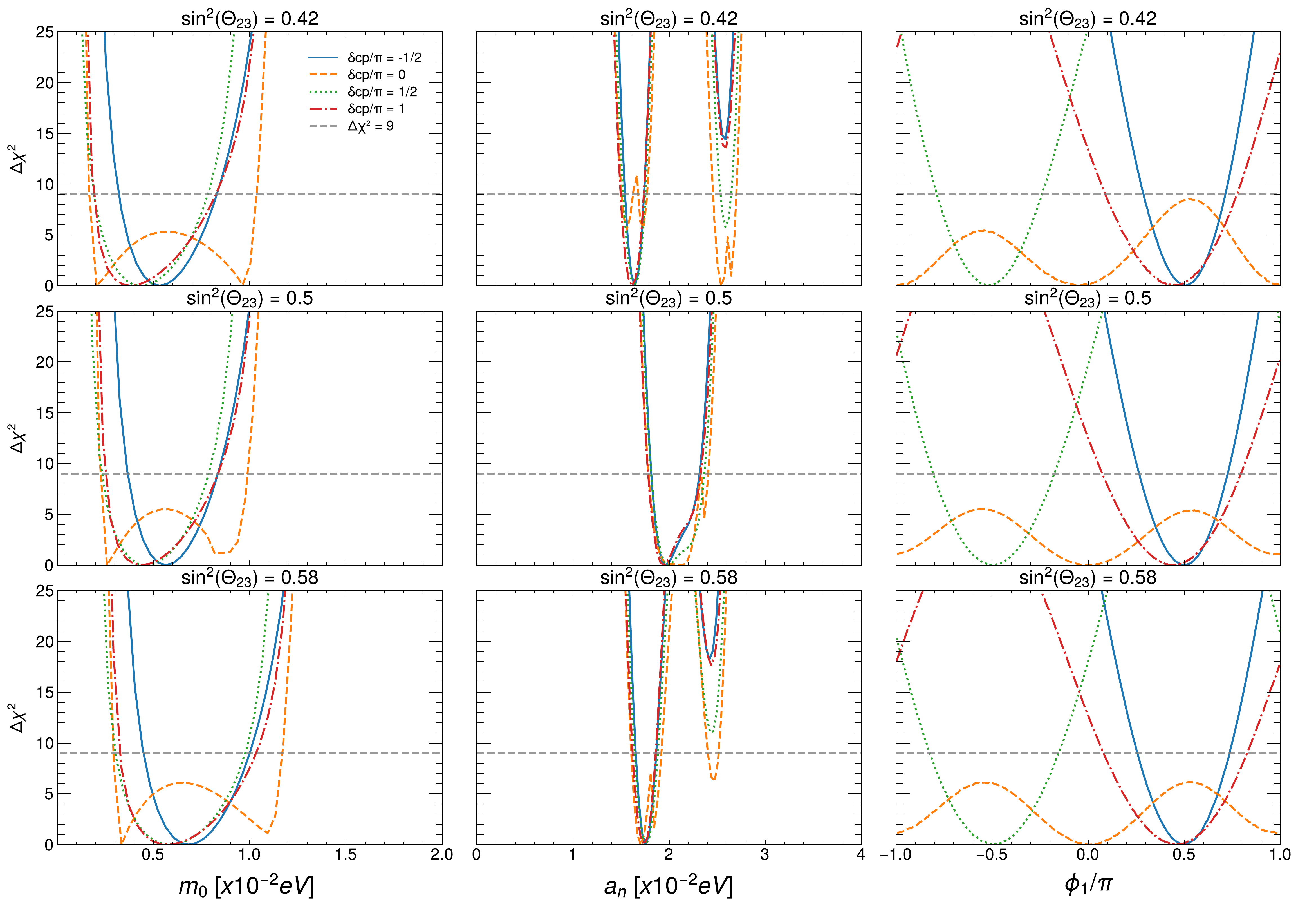}
    \caption{$\Delta \chi^2$ profiles for each one of the three mixing parameters $\vec{\lambda}=\{m_0,a_n,\phi_1\}$ that parametrize the mixing obtained from the $\text{RRR}_4$-form, after marginalization over the other two not shown parameters. The upper, middle, and lower row panels correspond to results for the input values $\sin^2{\theta}_{23}=\{0.42, 0.5, 0.58\}$, respectively. The lines correspond to results for the input values $\delta_{\text{CP}}/\pi=\{-0.5,0,0.5,1\}$, respectively. The horizontal dashed line corresponds to a $3\sigma$ of C.L for d.o.f. More details are given in the text and in Table~\ref{tab:RRR44interval}. }
    \label{fig:RRR44layout}
\end{figure}

In Fig.~\ref{fig:RRR44layout} we present the $\Delta \chi^2$ profiles for each one of the mixing parameters $\vec{\lambda}$ in terms of input values of the atmospheric mixing angle $\sin^2 \theta_{23}=\{0.42,0.5,0.58\}$ and $\delta_{\text{CP}}/\pi=\{-0.5,0,0.5,1\}$. In all the cases, the sensitivities to $m_0$ and $a_n$ parameters are bounded at a high-confidence level with some dependence of the $\delta_{\text{CP}}$ input value. More importantly, DUNE is sensitive to the phase $\phi_1$, although with an important dependence on the assumed input value for $\delta_{\text{CP}}$. From the left-vertical panels, one can see that DUNE has a reduced sensitivity to the $m_0$ parameter in comparison to the constraint, as seen in Fig.~\ref{fig:RRR44profiles} or by comparing the values in the second column in Table~\ref{tab:RRR44interval} with Table~\ref{tab:rrr4fit} calculated at the same confidence ($3\sigma$ C.L for one d.of.). In the case of $a_n$, in the middle vertical panels, DUNE is sensitive to approximately the same range of values in Table~\ref{tab:rrr4fit} as shown in the third column in Table~\ref{tab:RRR44interval} with a kind of separation in two solutions, which is correlated with the octant assumed for $\sin^2 \theta_{23}$. Finally, DUNE is sensible to the phase $\phi_1$, although the values are shifted approximately by $\pm \pi/2$ for $\delta_{\text{CP}}\ne 0$, in comparison to the constraint in the right panel of Fig.~\ref{fig:RRR44profiles}. Only the case of the sensitivity for $\delta_{\text{CP}}=0$ is compatible with the constraint. Notice, however, that we have not included the constraints on $a_n$ and $m_0$ for the sensitivity analysis in DUNE. Instead, the sensitivities are obtained from a DUNE-only analysis, without external priors. Also, a strict comparison between both analyses is not possible since we are simulating DUNE events without statistical fluctuations. The main purpose is to show that DUNE is sensitive to all the $\vec{\lambda}$ parameters, particularly to the $CP$-violating phase $\phi_1$.

Before closing this section, it is worth commenting on the values of the $\chi^2$ function at the best-fit point $\chi^2_{\text{min}}$ shown in the last column in Table~\ref{tab:RRR44interval}. For any of the $\sin^2 \theta_{23}$ values, the minimum value is obtained for $\delta_{\text{CP}}=0$, showing agreement between the assumed fake data set and the events calculated using the mixing hypothesis. In all the remaining cases, a `tension' showing different level of disagreement between the assumed fake data and the mixing hypothesis results. Notice, however, that the number of degrees of freedom is larger than the highest $\chi^2_{\text{min}}$ value (81.5), given the number of bins in the DUNE simulation ($80$ bins in neutrino energy reconstructed $E_{\text{rec}}$, where $0\le E_{\text{rec}} \le10~\text{GeV}$, per oscillation channel). Thus, from a `goodness-of-fit test' criteria, it is not possible to ruled out any of the hypothesis, specially because we are simulating the data without statistical fluctuations. Again, this analysis shows that DUNE is sensitive to all the $\vec{\lambda}$ parameters without conclusive statements on the rejection of any of the forms presented along this letter. We need to wait for DUNE data to repeat the analysis proposed here or to add statistical fluctuations to the fake data which is out of the scope of the present letter.

In the next section, we introduce one of the forms where the DUNE sensitivity range of the physical phases is better than the constraint region at the same C.L.

\begin{table}[t]
\begin{tabular}{|c c | c c c | c|}
        \hline
           $\sin^2(\theta_{23})$ &$ \delta_{\text{CP}}/\pi$ & $m_0\ (\times 10^{-2}~\text{eV})$ & $a_n$ $(\times 10^{-2}~\text{eV})$ & $\phi_1/\pi$ & $\chi^2_{\text{min}}$ \\ 
           \hline \hline
           0.42 & -0.5& [0.3, 0.8] &  [1.5, 1.7]                 & [0.3  0.7]   & 53.3 \\
           0.42 & 0.0 & [0.2, 1.1] & $[1.5,1.6]\cup [2.4, 2.7]$ &  [-1.0, 1.0]  & 0 \\
           0.42 & 0.5 & [0.2, 0.8] & $[1.5, 1.7]\cup [2.5, 2.6]$ & [-0.7, -0.2]  & 30.3 \\
           0.42 & 1.0 & [0.2, 0.8] &  [1.5, 1.7]                 & [0.0, 0.8]   & 76.6 \\
           \hline            
           0.50 & -0.5& [0.5, 0.8] &  [1.8, 2.3]                 & [0.3,  0.7]   & 55.1 \\
           0.50 & 0.0 & [0.3, 1.0] &  [1.8, 2.4]                 & [-1.0, 1.0]& 0 \\
           0.50 & 0.5 & [0.3, 0.8] &  [1.8, 2.4]                 & [-0.8, -0.2] & 30.1 \\
           0.50 & 1.0 & [0.3, 0.8] &  [1.8, 2.3]                 & [0.1, 0.8]   & 81.5 \\
           \hline      
           0.58 & -0.5& $[0.5, 1.0]$ &  [1.6, 1.9]                 & [0.3, 0.7]  & 48.0 \\ 
           0.58 & 0.0 & $[0.3, 1.2]$ &  $[1.7, 1.8]\cup [2.4, 2.5]$ & [-1.0, 1.0] & 0 \\
           0.58 & 0.5 & $[0.3, 1.0]$ &  [1.6, 1.9]                 & [-0.8, -0.2] & 26.5 \\
           0.58 & 1.0 & [0.9, 1.1] &    [1.6, 1.9]                 & [0.1, 0.8]  & 75.5 \\ [1ex]
           \hline 
           
\end{tabular}
\caption{\label{tab:RRR44interval} Parameter ranges at $3\,\sigma$ of C.L for 1 d.o.f. from the sensitivity results in Fig.~\ref{fig:RRR44layout} for each one of the three mixing parameters $\vec{\lambda}=\{m_0,a_n,\phi_1\}$ that parametrize the mixing obtained from the $\text{RRR}_4$-form, for the different pair of input values of $\sin^2{\theta}_{23}$ and $\delta_{\text{CP}}$, as quoted in the first two columns of the table. The $\chi^2$ values at the bfp $\chi^2_{\text{min}}$, are also included in the last column. See text for more details.}
\end{table}

\subsection{Results for the $\text{T}_1$-form}
In this section we report on the DUNE sensitivity to the mixing predicted by the form $\text{T}_1$-form in Table~\ref{tab:rrr}. We have followed the same procedure as the one applied in the previous section.

\begin{figure}[t]
    \centering
    \includegraphics[width=\textwidth]{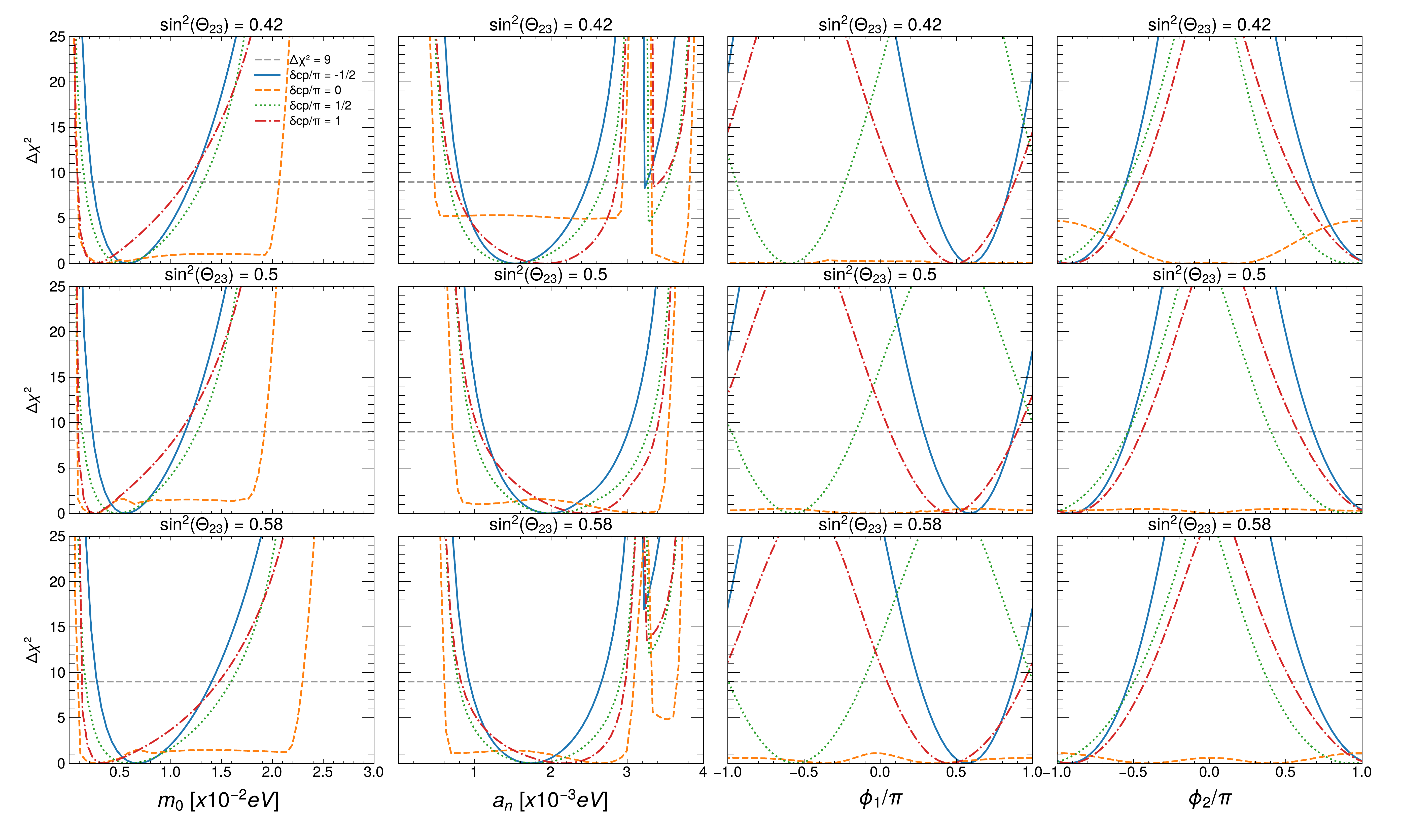}
    \caption{$\Delta \chi^2$ profiles for each one of the four mixing parameters $\vec{\lambda}=\{m_0, a_n, \phi_1, \phi_2\}$ that parametrize the mixing obtained from the $\text{T}_1$-form, after marginalization over the other three not shown parameters. The upper, middle, and lower row panels correspond to results for the input values $\sin^2{\theta}_{23}=\{0.42, 0.5, 0.58\}$, respectively. The lines correspond to results for the input values $\delta_{\text{CP}}/\pi=\{-0.5,0,0.5,1\}$, respectively. The horizontal dashed line corresponds to a $3\sigma$ of C.L for d.o.f. More details are in the text and in Table~\ref{tab:T11interval}. }
    \label{fig:T11layout}
\end{figure}

In Fig.~\ref{fig:T11layout} we present the $\Delta \chi^2$ profiles for each one of the four mixing parameters $\vec{\lambda}$ in terms of input values of the atmospheric mixing angle $\sin^2 \theta_{23}$ and $\delta_{CP}$ assuming normal neutrino mass ordering. Similar to $\text{RRR}_4$-form, the sensitivities to $m_0$ and $a_n$ parameters are bounded at a high-confidence level almost independent of the $\delta_{\text{CP}}$ input value. From the left-vertical panels, one can see that DUNE has a reduced sensitivity to the $m_0$ parameter in comparison to the constraint in Fig.~\ref{fig:T11profiles} and in Table~\ref{tab:t1fit}. In the case of $a_n$, in the middle vertical panels, DUNE is sensitive to approximately the same range of values in Table~\ref{tab:rrr4fit} with a kind of separation in two solutions, which is correlated with the octant assumed for $\sin^2 \theta_{23}$. Different to $\text{RRR}_4$-form, in this form, DUNE sensitivity to the $\phi$ phases is compatible with the constraints from the lower panels in Fig.~\ref{fig:T11profiles}. More importantly, the sensitivity range at $3\sigma$ C.L for 1 d.o.f., in Table~\ref{tab:T11interval}, is smaller than the constraint at the same confidence in Table~\ref{tab:t1fit}. It is then expected that DUNE is sensitive to the $CP$ violation, encoded in the $CP$-violating phases, within the analyzed forms. This is the subject of the next section.

\begin{table}[t]
        \begin{tabular}{|c c| c c c c |c|}
        
        \hline
           $\sin^2(\theta_{23})$ &$ \delta_{cp}/\pi $& $ m_0\ (\times 10^{-2}~\text{eV})$ & $a_n$ $(\times 10^{-2}~\text{eV})$ & $\phi_1/\pi$ &   $\phi_2/\pi$ & $\chi^2_{\text{min}}$ \\ [0.5ex] 
        \hline \hline
        0.42 &  -0.5 &  [0.3, 1.4] &  [0.9, 2.5] &  $[-1.0,  -0.6]\cup [0.7,  1.0]$ &  $[-1.0,  -0.5]\cup [0.6, 1.0]$ & 19.4  \\
        0.42 &   0.0 &  [0.1, 2.2] &  $[0.5, 3.0] \cup [3.1, 3.8]$ &                 [-1.0,  1.0] & [-1.0,  1.0] &  5.0 \\
        0.42 &   0.5 &  [0.3, 1.3]  & $[0.7, 2.8] \cup [3.1, 3.3]$ &                 [-0.9,  -2.4] & $[-1.0,  -0.6]\cup [0.4,  1.0]$ & 13.0 \\
        0.42 &     1 &  [0.2, 1.2]  & [0.7, 2.8] &                 [0.1,  8.72] &  $[-1.0,  -0.5]\cup [0.5,  1.0]$ &  59.4\\
        \hline                
        0.5 &  -0.5 &   [0.4, 1.3] & [1.2, 3.0]  &               [0.2 ,  0.9] &  [-0.5,  0.7] &  17.7 \\
        0.5 &   0.0 &   [0.1, 2.0] & [0.6, 3.4]  &                [-1.0,  1.0] &  [-1.0,  1.0] &  1.04 \\
        0.5 &   0.5 &   [0.1, 1.3] & [1.0, 3.3]  &                [-1.0,  -0.1] & [-0.5,  0.4] &  11.9 \\
        0.5 &  1 &      [0.1, 1.2] & [1.0, 3.2]   &                [0.0,  0.8] &   [-0.4,  0.6] &  62.3 \\
        \hline                
         0.58 &  -0.5 & [0.4, 1.5] & [1.0, 2.5]  &               [0.3 ,  0.9]  & [-0.5,  0.6] &  14.7\\
         0.58 &     0 & [0.1, 2.4] & $[0.6, 3.0]\cup[3.3, 3.7]$  &               [-1.0,  1.0] &  [-1.0,  1.0] &  0\\
         0.58 &   0.5 & [0.2, 1.6] & [0.9, 2.9] &                [-1.0,  0.0]&   [-0.5,  0.4 ] & 10.5 \\ 
         0.58 &     1 & [0.2, 1.5] & [0.9, 2.9]    &             [0.1,  0.9] &   [-0.4,  0.5] &  58.5\\  [1ex]
        \hline
           
        \end{tabular}
        \caption{\label{tab:T11interval} Parameter ranges at $3\,\sigma$ of C.L for 1 d.o.f. from the sensitivity results in Fig.~\ref{fig:T11layout} for each one of the four mixing parameters $\vec{\lambda}=\{m_0,a_n,\phi_1, \phi_2\}$ that parametrize the mixing obtained from the $\text{T}_1$-form, for the different pair of input values of $\sin^2{\theta}_{23}$ and $\delta_{\text{CP}}$, as quoted in the first two columns of the table. The $\chi^2$ values at the bfp $\chi^2_{\text{min}}$, are also included in the last column. See text for more details.}
        \end{table}

\subsection{DUNE sensitivity to the $CP$-violating phases}
The main goal of this letter is to quantify the DUNE sensitivity to amount of $CP$ violation within each one of the forms, given the present constraints on the $m_0$ and $a_n$ parameters. As shown in Figs~\ref{fig:RRR44profiles}, \ref{fig:T11profiles}, \ref{fig:RRR11profiles}, and \ref{fig:RRR33profiles}, both $m_0$ and $a_n$ are constrained by the current values of the neutrino oscillation mixing angles. In addition, DUNE is sensitive to the CP-violating phases present in the forms as shown in Figs~\ref{fig:RRR44layout}, \ref{fig:T11layout}, \ref{fig:RRR11layout}, and \ref{fig:RRR33layout}. Therefore, we can use the results presented up to now, to quantify the sensitivity to the amount of $CP$ violation within the forms, which can be quantified through the Jarlskog invariant. 

In the forms with three degrees of freedom ($\text{RRR}_1$ and $\text{RRR}_4$) the Jarlskog invariant has a one-to-one correspondence with the physical phase $\phi_1$ [see Eq.~(\ref{eq:correspondence})], as happens with the Dirac $CP$ phase in the PDG parametrization of the lepton mixing. However, In the forms with four degrees of freedom, $\text{T}_1$ and $\text{RRR}_3$, the Jarlskog invariant depends on the two $\phi$ phases in a nontrivial way, as shown in Figs~\ref{fig:T11Jcp} and \ref{fig:TRRR33Jcp}, respectively. Given the extra freedom in this models, and without lack of generality, we have opted for fixing the phase $\phi_2$ to a constant value. The criteria to choose a convenient value for this phase is based on the Jarlskog contours in Figs~\ref{fig:T11Jcp} and \ref{fig:TRRR33Jcp}. Basically, one can see that the Jarlskog reaches a maximum/minimum value when $\phi_1=\pm 0.5\pi$ for values of $\phi_2 \approx \pm \pi$, which are both equivalent. Thus, $\phi_1$ is a natural choice for accounting for the Dirac $CP$ phase, which covers all physical values including zero. Therefore, for the rest of this section we have defined $\phi_2\equiv \pi$. Notice that this choice is in agreement with the allowed values in Figs~\ref{fig:T11profiles} and \ref{fig:RRR33profiles}.

To quantify the DUNE sensitivity to the $CP$-violating phase we have used the  $\chi^2$ function in Eq.~(\ref{eq:chi2}) replacing $\vec{\theta}_0 \to \vec{\lambda}_0\equiv \vec{\lambda}^{\text{True}}$ and testing for $CP$ conserving values of the phase $\phi_1^{\text{test}}=\{-\pi,0,\pi\}$. Notice that $\vec{\lambda}^{\text{True}}=\{m_0^{\text{True}}, a_n^{\text{True}}, \phi_1^{\text{True}}\}$. The values for the true parameters $m_0$ and $a_n$ are fixed to the best-fit values in Tables~\ref{tab:rrr4fit}, \ref{tab:t1fit}, \ref{tab:rrr1fit} and \ref{tab:rrr3fit} which are slightly different for each one of the forms. In the minimization over $m_0^{\text{Test}}$ and $a_n^{\text{Test}}$, the $\chi^2$ function is penalized with the $\Delta \chi^2$ profiles,  $\Delta \chi^2(m_0)$ and $\Delta \chi^2(a_n)$, using the results in Figs~\ref{fig:RRR44profiles}, \ref{fig:T11profiles}, \ref{fig:RRR11profiles}, and \ref{fig:RRR33profiles}. This is the correct way of proceeding since, in general, the $\Delta \chi^2$ profiles of this parameters are not Gaussian.

\begin{figure}[t]
    \centering
    \includegraphics[width=\textwidth]{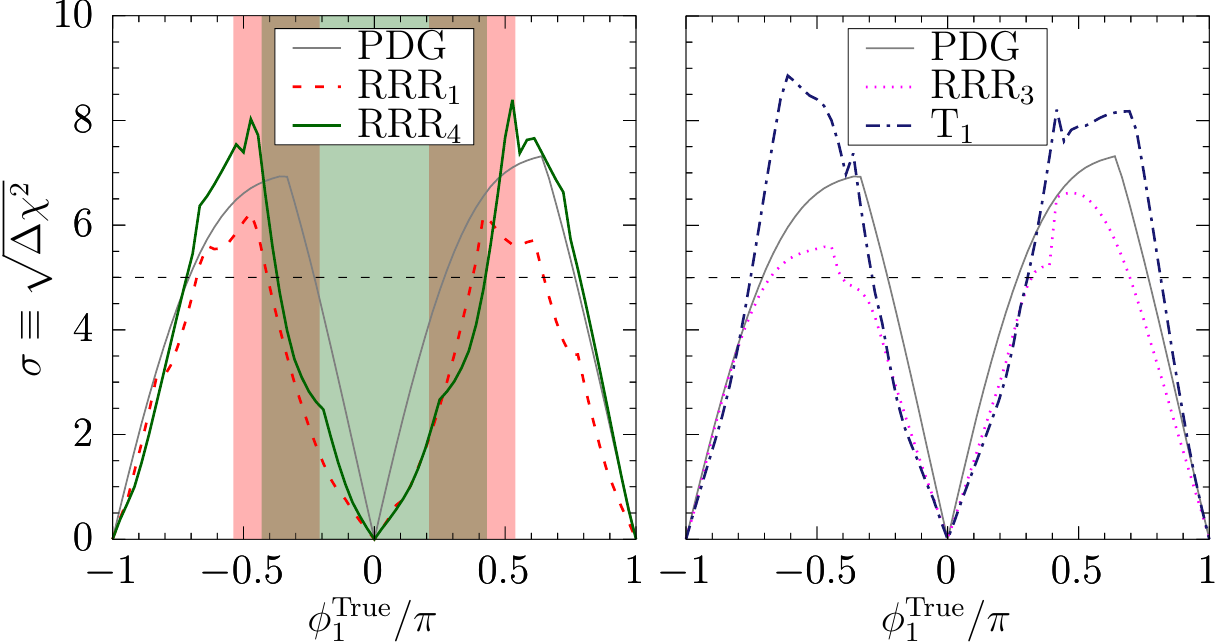}
    \caption{DUNE sensitivity to the $CP$-violating phase $\phi_1$. In the left (right) panel the sensitivity obtained for the $\text{RRR}_1$ and $\text{RRR}_4$ ($\text{RRR}_3$ and $\text{T}_1$) forms in the full and the dashed lines (dotted and dot-dashed lines), respectively. In the case of the $\text{RRR}_1$ and $\text{RRR}_4$-forms, the allowed ranges at $3\sigma$ of C.L. (see Table~\ref{tab:rrr1fit} and Table~\ref{tab:rrr4fit}) have also been included in the left panel as colored regions in red and green, respectively. In the case of the textures in the right panel, we have fixed $\phi_2=\pi$ so that there is a one-to-one correspondence between $\phi_1$ and the Jarlskog invariant, see text for details. The sensitivitity to the Dirac $CP$ phase in the PDG parametrization of the lepton mixing is also included as a reference in both panels as the full line in gray (light) color. The horizontal line corresponds to a significance level of $5\sigma$.}
    \label{fig:CPsensitivity}
\end{figure}

Figure~\ref{fig:CPsensitivity} shows the DUNE sensitivity to the $CP$-violating phase $\phi_1$, which has a one-to-one correspondence with the Jarlskog invariant in the case of $\text{RRR}_1$ (dashed line) and $\text{RRR}_4$ (full line in dark color) forms shown in the left panel. After fixing $\phi_2=\pi$ this one-to-one correspondence is also present in the case of the $\text{T}_1$ (dotted-dashed line) and $\text{RRR}_3$ (dotted line) forms, and the sensitivity to the Dirac $CP$ phase $\phi_1$ is shown in the right panel. Interestingly enough, the DUNE sensitivity to the $CP$-violating phase $\phi_1$, at $5\sigma$, for all the forms, is comparable with the sensitivity to the $\delta_{\text{CP}}$ phase in the PDG parametrization (full line in light color) within the corresponding $\phi_1$-allowed range. In the case of the $\text{RRR}_1$ and $\text{RRR}_4$-forms, the allowed ranges at $3\sigma$ of C.L. (see Table~\ref{tab:rrr1fit} and Table~\ref{tab:rrr4fit}) have also been included in the left panel as colored regions in red and green, respectively.  In particular, at a significance of $5\sigma$ the $CP$ coverage for both $CP$ phases in the $\text{T}_1$-form and the PDG parametrization, is $49\%$. In the case of maximal $CP$ violation, DUNE sensitivity is higher for the $\text{T}_1$ and $\text{RRR}_4$-forms than for the PDG one. All the studied forms are interesting proposals from the model building point of view. The $\text{RRR}_1$-form, with three degrees of freedom, is fully compatible with the current neutrino oscillation phenomenology including a prediction for the Dirac $CP$ phase compatible with maximal $CP$ violation (see Table~\ref{tab:rrr1fit}). Finally, the form $\text{T}_1$-form have enough freedom to accommodate the neutrino oscillation phenomenology, with a similar $CP$ coverage as the $\delta_{\text{CP}}$ phase in the PDG parametrization, and with higher significance for nearly maximal values compatible with $CP$ violation.

\section{\label{sec:summary} Summary and conclusions}
%
In the model considered here (Dirac neutrinos), and using the polar decomposition theorem of the matrix algebra, the mass matrices can be finally written in terms of Hermitian matrices implying that a zero in the position $(i,j)$ is the same zero in the position $(j,i)$ and therefore it is counted only once in the texture. Since texture zeros can be arbitrarily placed in any of the elements of a mass matrix, this leads to a large number of possibilities. In this work we have considered the five texture zeros inspired by Ramond, Robert, and Roos forms in the quark sector, assuming they can be applied to the lepton sector, which seems a reasonable assumption for hierarchical neutrinos. This hypothesis has been tested and only three out of the five RRR-forms are nonequivalent, given the texture zeros assignment in the neutral lepton sector for this three forms.

Texture zeros provide a convenient frame in a theory that lacks flavor structure, and becomes useful accomplishing two goals. The first one, diminishing the mathematical parameters in the models, as has been shown in the case of five texture zeros in an extension of the SM with three right-handed singlet Dirac neutrinos. The second one, a prediction of the lepton mixing matrix obtained from the rotation matrices that diagonalize the forms assumed for the charged and neutral sectors mass matrices, thus providing an alternative to the PDG parametrization. We have developed a general procedure to obtain the lepton mixing in terms of four free parameters (the absolute neutrino mass $m_0$, $a_n$, and the two phases, $\phi_1$ and  $\phi_2$), which can be determined from the current values of the neutrino oscillation mixing angles. Notice, however, that the forms $\text{TRRR}_1$ and $\text{TRRR}_4$ depend on only three free parameters, while compatible with current neutrino phenomenology, showing the convenience of the assumed texture zeros.

All the forms in Table~\ref{tab:rrr} predict a neutrino mixing compatible with the current neutrino oscillation phenomenology including the possibility of the $CP$-symmetry violation encoded in a physical phase. We have shown that DUNE sensitivity to this Dirac $CP$ phase is comparable to the sensitivity to $\delta_{\text{CP}}$ of the PDG parametrization. 

The procedure followed in the analysis of the predicted mixing serves as a criteria to determine nonequivalent forms from phenomenological grounds. In all the forms studied here, a prediction of the absolute neutrino mass, and its range, was found in agreement with the current upper limits. As a consequence, the theory is richer and falsifiable, increasing the options to provide an insight of the flavor structure of the lepton sector.

\section*{Acknowledgments}
We would like to thank ITM because this research was partly supported by the ‘Centro de Laboratorios de investigación parque i-ITM, P20246 project’. A.~T. and D.~V.~F. also thank ‘Vicerrectoría de CyT - UdeMedellin’ for all the support during the development of this work. 

\clearpage

\appendix
\section{RESULTS FOR THE $\text{RRR}_{1}$-FORM}
\label{sec:app1}
\subsection{Analytical and numerical analysis}\label{subsec:app1a}
The $\text{RRR}_{1}$-form was defined in Table~\ref{tab:rrr}, and by applying the procedure discussed in the main text, the following analytical results are obtained:

\begin{equation}
\begin{split}
|b_n| &=\sqrt{(m_1 m_2 m_3)/a_n},\\ c_n & =-a_n + m_1 - m_2 + m_3,\\ |d_n| &=\sqrt{-(a_n - m_1) (a_n + m_2) (a_n - m_3)/a_n}.\\
\end{split}
\end{equation}

The following orthogonal rotation matrix diagonalize the mass matrix assumed for the neutral lepton sector:

\begin{widetext} 
\begin{equation}
R_n=\left(
\begin{array}{ccc}
 -\sqrt{\frac{m_2 m_3(a_n - m_1)}{a_n (m_1 + m_2) (-m_1 + m_3)}} & -\sqrt{\frac{m_1(a_n - m_1) }{(m_1 + m_2) (-m_1 + m_3)}} & \sqrt{\frac{m_1 (a_n + m_2) (a_n - m_3)}{a_n (m_1 + m_2) (m_1 - m_3)}} \\
 \sqrt{\frac{m_1 m_3(a_n + m_2)}{a_n (m_1 + m_2) (m_2 + m_3)}} & -\sqrt{\frac{m_2 (a_n + m_2)}{(m_1 + m_2) (m_2 + m_3)}} & \sqrt{\frac{-m_2(a_n - m_1)(a_n - m_3)}{a_n (m_1 + m_2) (m_2 + m_3)}} \\
 \sqrt{\frac{m_1 m_2 (a_n - m_3)}{a_n (m_1 - m_3) (m_2 + m_3)}} & \sqrt{\frac{m_3 (-a_n + m_3)}{(-m_1 + m_3) (m_2 + m_3)}} & \sqrt{\frac{m_3(a_n - m_1) (a_n + m_2)}{a_n (-m_1 + m_3) (m_2 + m_3)}} \\
\end{array}
\right).
\label{rn}
\end{equation}
\end{widetext}

The results for the charged sector are equal to the ones obtained for $\text{RRR}_4$-form. To ensure all elements in the $R_n$ matrix are real, the mathematical restriction $m_1 < a_n < m_3$ must be satisfied. Thus, given a constraint on $m_0$, the $a_n$ is bounded by the neutrino masses in Eq.~(\ref{eq:numasses}). Independent of the current upper bounds on $m_0$, the largest allowed range is obtained when $m_0\ll 1$, and therefore, $0<a_n<\sqrt{\Delta m^2_{31}}$ or numerically $0<a_n/\text{eV} < 5.1\times 10^{-2}$ using the values in Table~\ref{tab:oscparams}.

\begin{figure}[t]
    \centering
    \includegraphics[width=\textwidth]{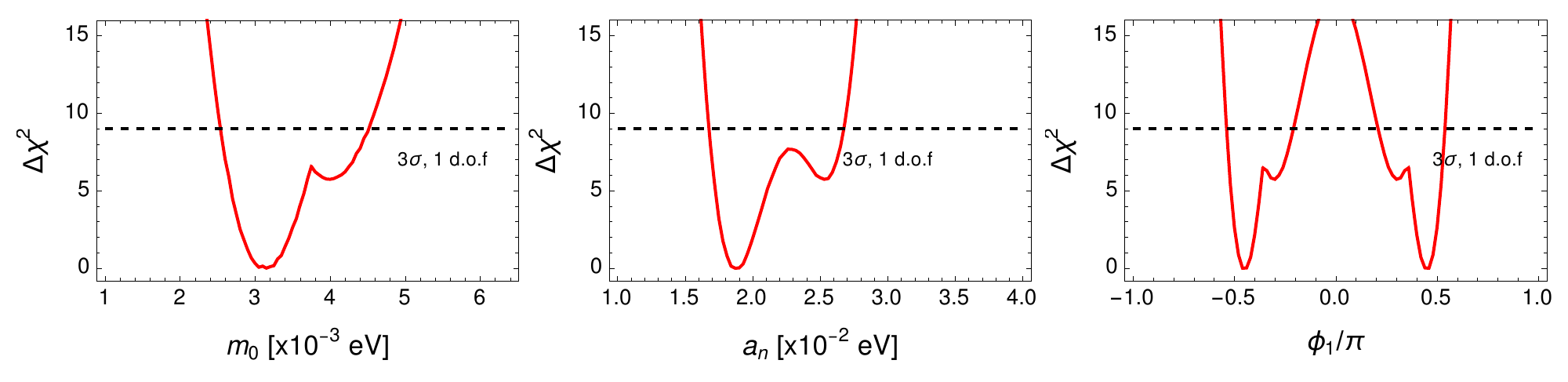}
    \caption{$\Delta \chi^2$ profiles for each one of the three mixing parameters $\vec{\lambda}=\{m_0,a_n,\phi_1\}$ (shown in left, middle and right panels, respectively) that parametrize the lepton mixing obtained from the $\text{RRR}_1$-form, after minimizing over the two parameters not shown in each panel. The horizontal line corresponds the allowed range at $3\sigma$ of C.L for $1\,\text{d.o.f.}$ whose values are reported in Table~\ref{tab:rrr1fit}. }
    \label{fig:RRR11profiles}
\end{figure}

The mixing is independent of $\phi_2$ in this texture, therefore the numerical results are given in terms of three parameters $\vec{\lambda}=\{m_0,a_n,\phi_1\}$. The Fig.~\ref{fig:RRR11profiles} shows the $\Delta \chi^2$ profiles for the three parameters, showing all the parameters are bounded. The limits appear in Table~\ref{tab:rrr1fit}.

The $\phi_1$ phase is the only source of $CP$ violation in the $\text{RRR}_1$-form. From the right panel in Fig.~\ref{fig:RRR11profiles} one can see that $CP$ conserving values are excluded at $3\sigma$. From Table~\ref{tab:rrr1fit}, the phase is restricted to the range $[-0.5\pi, -0.2\pi] \cup [0.2\pi, 0.5\pi]$ at $3\sigma$ of C.L for $1\,\text{d.o.f.}$, implying a restriction on the Jarlskog invariant, $J_{\text{CP}}/100 \in [-1.3,\ -0.7] \cup [0.7, 1.3]$ at the same C.L.

\begin{table}[t]
\begin{center}
\begin{tabular}{| c | c | c | }
\hline
Parameter & Best-fit & $3\sigma$ range \\ \hline \hline
$m_0$ ($\times 10^{-3}\text{eV}$) & $3.2$ & $[2.5, 4.5]$ \\ \hline
$a_n$ ($\times 10^{-2}\text{eV}$) & $1.9$ & $[1.7, 2.7]$ \\ \hline
$\phi_1/\pi$ & $\pm 0.5$ & $[-0.5, -0.2] \cup \,[0.2, 0.5]$  \\ \hline
$\phi_2/\pi$ & Independent & - \\ \hline 
\end{tabular}
\caption{\label{tab:rrr1fit} Best-fit parameters (second column) and three sigma allowed range for $1\,\text{d.o.f.}$ (third column) for each of the parameters of the $\text{RRR}_1$-form (first column) obtained from the minimization of Eq.~(\ref{eq:chi2NOexp}). The $\chi^2$ value at the minimum is $\chi^2_{\text{min.}}=0.12$.} 
\end{center}
\end{table}

\subsection{DUNE sensitivity to the mixing parameters}
\label{subsec:app1b}

\begin{figure}[t]
    \centering
    \includegraphics[width=\textwidth]{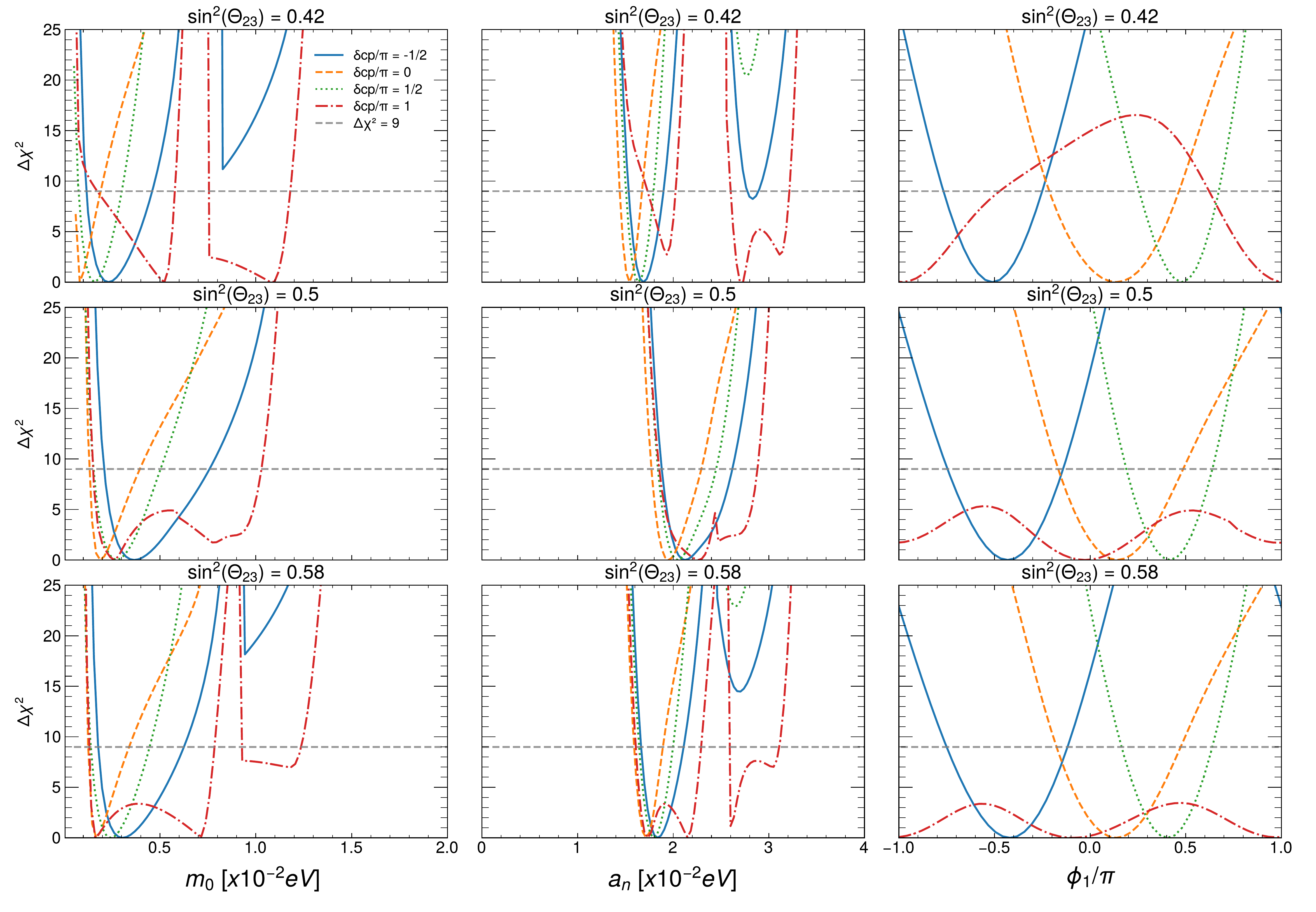}
    \caption{$\Delta \chi^2$ profiles for each one of the three mixing parameters $\vec{\lambda}=\{m_0,a_n,\phi_1\}$ that parametrize the mixing obtained from the $\text{RRR}_1$-form, after marginalization over the other two not shown parameters. The upper, middle, and lower row panels correspond to results for the input values $\sin^2{\theta}_{23}=\{0.42, 0.5, 0.58\}$, respectively. The lines correspond to results for the input values $\delta_{\text{CP}}/\pi=\{-0.5,0,0.5,1\}$, respectively. The horizontal dashed line corresponds to a $3\sigma$ of C.L for d.o.f. More details are in the text and in Table~\ref{tab:RRR11interval}. }
    \label{fig:RRR11layout}
\end{figure}

In this case we use simulated events at the DUNE FD to perform the analysis described in the main text (see Sec.~\ref{sec:expfit}). In Fig.~\ref{fig:RRR11layout} we show the $\Delta \chi^2$ profiles for each one of the $\vec{\lambda}$ parameters in terms of input values of the atmospheric mixing angle $\sin^2 \theta_{23}=\{0.42,0.5,0.58\}$ and $\delta_{\text{CP}}/\pi=\{-0.5,0,0.5,1\}$, while the remaining standard oscillation parameters were fixed to their best-fit values in Table~\ref{tab:oscparams}. In Table~\ref{tab:RRR11interval} we show the sensitivity ranges at $3\,\sigma$ of C.L for 1 d.o.f. from the results in Fig.~\ref{fig:RRR11layout}, for each one of the three parameters.

\begin{table}[t]
        \begin{tabular}{|c c| c c c |c|}
        
            \hline
               $\sin^2(\theta_{23})$ &$ \delta_{\text{CP}}/\pi $& $ m_0\ (\times 10^{-2}~\text{eV})$ & $a_n$ $(\times 10^{-2}~\text{eV})$ & $\phi_1/\pi$ & $\chi^2_{\text{min.}}$ \\ [0.5ex] 
            \hline \hline
            0.42 &  -0.5 &   [0.2,  0.5]               &  [1.5,  2.0]      &  [-0.9,  -0.1]     &  25.5\\
            0.42 &     0 &   [0.0,  0.2]               &  [1.5, 1.8]         &   [-0.1,  0.4]  &    0\\
            0.42 &   0.5 &   [0.2,  0.4]               &  [1.4,  1.7]                     &    [0.2,  0.8] &    89.5\\
            0.42 &     1 &   $[0.3,  0.6] \cup [0.7,  1.4]$ &  $[1.8,  2.1]\cup [2.8, 3.1]$                     &   [-1.0,  1.0] &    63.3\\
            \hline                                                  
             0.5 &  -0.5 &   [0.3,  0.7]                &  [1.9,   2.6]                    &  [-0.9,  -0.1] &  26.1\\
             0.5 &     0 &   [0.2,  0.4]                &  [1.8,  2.3]                     &   [-0.2,  0.5] & 71.9\\
             0.5 &   0.5 &   [0.2,  0.5]                &  [1.9,  2.4]                     &    [0.2,  0.7]  & 48.8 \\
             0.5 &     1 &   [0.2,  1.2]               &  [1.8,  2.9]                     &   [-1.0,  1.0]  & 0.9\\
            \hline                                  
            0.58 &  -0.5 &    [0.2,  0.6]                &  [1.7,  2.1]                     &  [-0.9,  -0.1] & 22.6\\
            0.58 &     0 &    [0.1,  0.3]                &  [1.6,  1.9]                     &   [-0.2,  0.4]  & 63.9\\
            0.58 &   0.5 &    [0.1,  0.4]              &  [1.6,  2.3]    &      [0.1,  0.6]   & 2.5 \\
            0.58 &     1 &    $[0.1,  0.7]\cup [0.9,  1.3]$                &  $[1.7,  2.0] \cup [2.8, 3.1]$                    &   [-1.0,  1.0]  & 42.4\\    [1ex]
            \hline                  
            \end{tabular}
        \caption{\label{tab:RRR11interval} Parameter ranges at $3\,\sigma$ of C.L for 1 d.o.f. from the sensitivity results in Fig.~\ref{fig:RRR11layout} for each one of the three mixing parameters $\vec{\lambda}=\{m_0,a_n,\phi_1\}$ that parametrize the mixing obtained from the $\text{RRR}_1$-form, for the different pair of input values of $\sin^2{\theta}_{23}$ and $\delta_{\text{CP}}$, as quoted in the first two columns of the table. The $\chi^2$ values at the bfp $\chi^2_{\text{min}}$ are also included in the last two columns. See main text for more details.}
        \end{table}

\section{RESULTS FOR THE $\text{RRR}_{3}$-FORM}
\label{sec:app2}
\subsection{Analytical and numerical analysis}\label{subsec:app2a}

The mixing from this texture, like the $\text{T}_1$-form discussed in the main text, is determined by four parameters: $\vec{\lambda}=\{m_0,a_n,\phi_1, \phi_2\}$. The rotation matrix $R_n$ corresponds to the one in the Appendix~\ref{sec:app1} because the same neutral mass matrix was assumed, while $R_l$ is the same as the one of the $\text{T}_1$-form. The mathematical restriction $m_1 < a_n < m_3$ must be satisfied implying $0<a_n<\sqrt{\Delta m^2_{31}}$ or numerically $0<a_n/\text{eV} < 5.1\times 10^{-2}$ using the values in Table~\ref{tab:oscparams}, as discussed in the main text.

\begin{figure}[t]
    \centering
    \includegraphics[width=\textwidth]{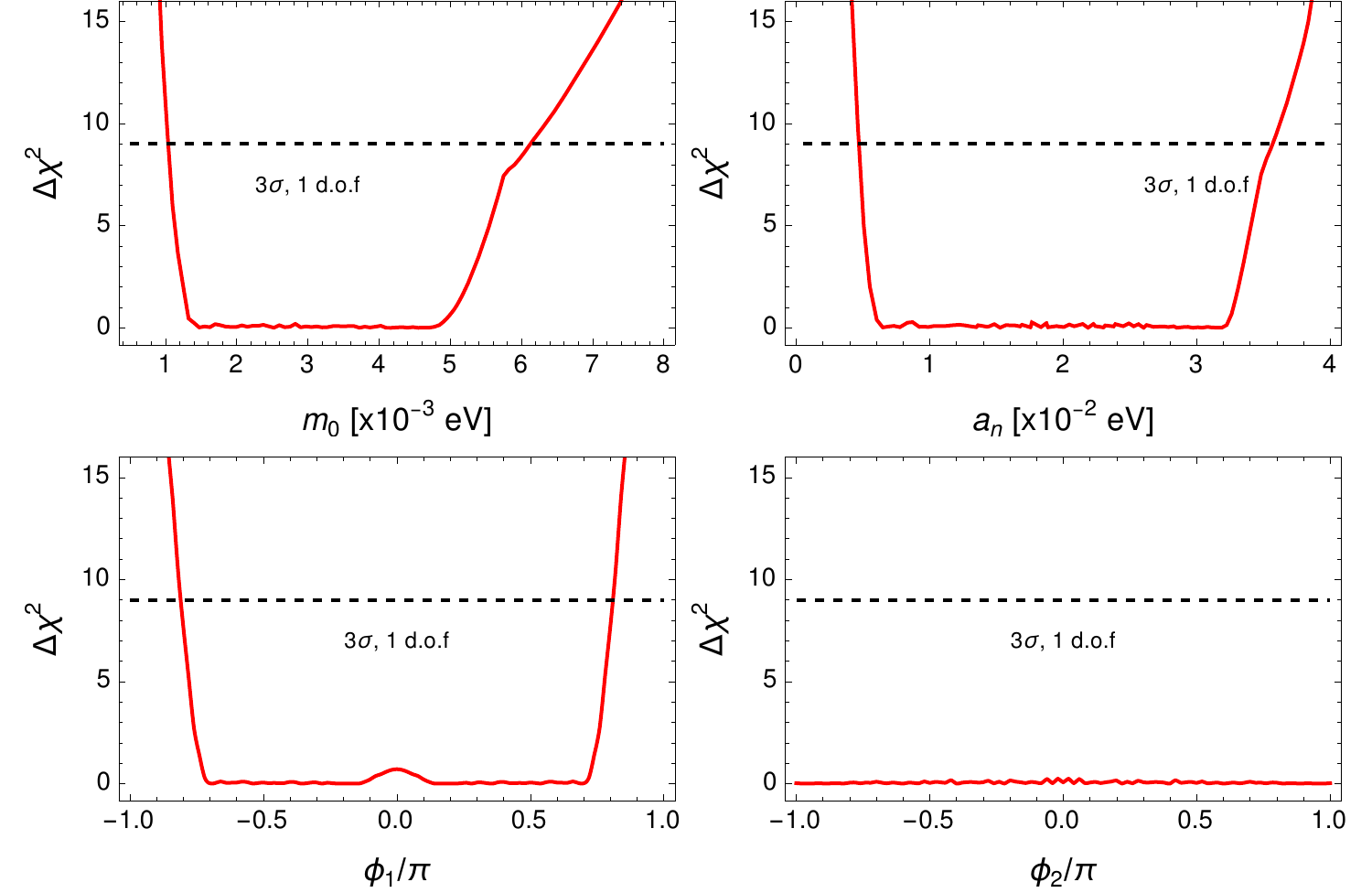}
    \caption{$\Delta \chi^2$ profiles for each one of the four mixing parameters $\vec{\lambda}=\{m_0,a_n,\phi_1, \phi_2\}$ (shown in the upper left, upper right, lower left and lower right panels, respectively) that parametrize the lepton mixing obtained from the $\text{RRR}_3$-form, after minimizing over the three parameters not shown in each panel. The horizontal line corresponds the allowed range at $3\sigma$ of C.L for $1\,\text{d.o.f.}$ whose values are reported in Table~\ref{tab:rrr3fit}. }
    \label{fig:RRR33profiles}
\end{figure}

\begin{figure}[t]
    \centering
    \includegraphics[width=0.7\textwidth]{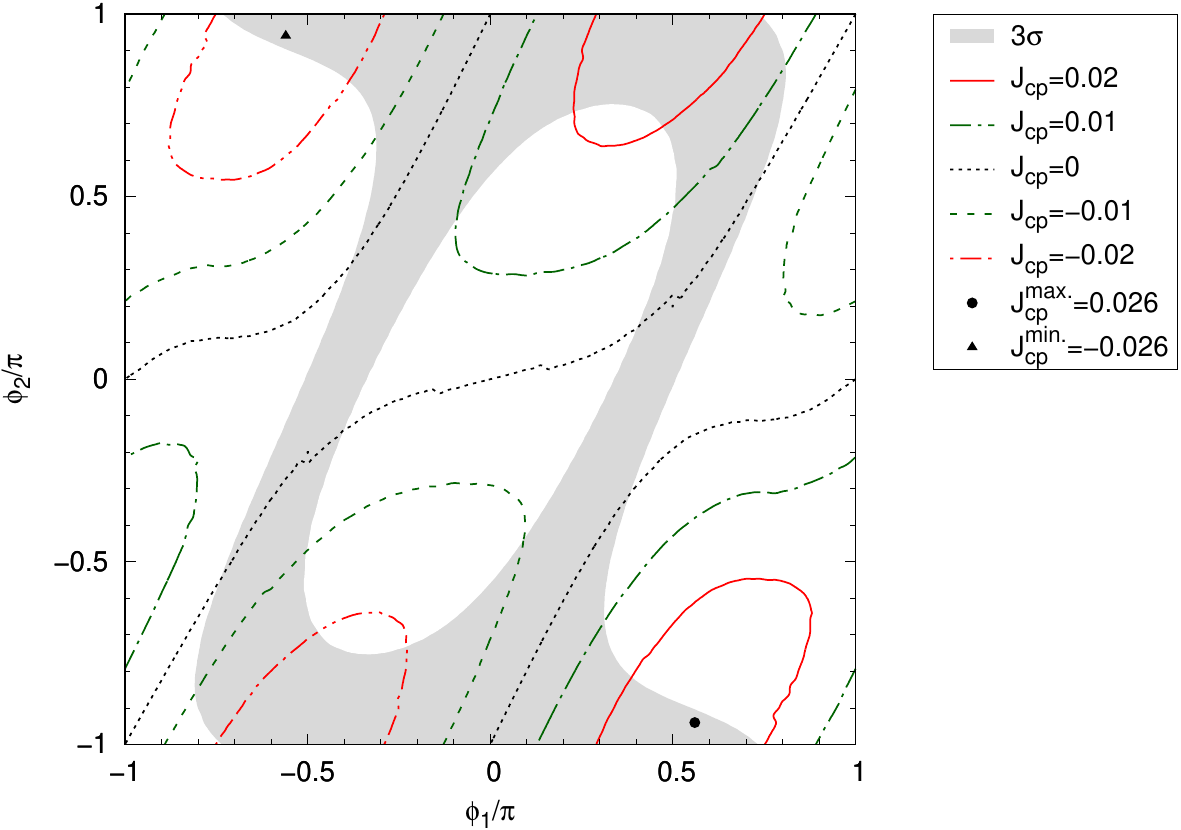}
    \caption{Jarlskog invariant in terms of the $\phi$ phases. The gray region, correspond to the allowed region at $3\sigma$ of C.L for $1\,\text{d.o.f.}$ after the marginalization of $m_0$ and $a_n$. The Jarlskog is evaluated at these $m_0$ and $a_n$. The lines correspond to some constant values of the Jarlskog. The maximum (minimum) values of the Jarlskog are also labeled with the point (triangle).}
    \label{fig:TRRR33Jcp}
\end{figure}

First, Fig.~\ref{fig:RRR33profiles} shows the $\Delta \chi^2$ profiles for each one of the four mixing parameters $\vec{\lambda}$ from the fit to the mixing angles. In Table~\ref{tab:rrr3fit} the best-fit parameters and three sigma allowed range for $1\,\text{d.o.f.}$ for each of the $\vec{\lambda}$ parameters are shown. In this form, the Jarlskog invariant depends of $\phi_1$ and $\phi_2$ phases (see Fig.~\ref{fig:TRRR33Jcp}). From the bottom right panel in Fig.~\ref{fig:RRR33profiles} one can see the $\phi_2$ phase is unconstrained at $3\sigma$ of C.L for $1\,\text{d.o.f.}$ and its profile is completely flat in its parameter space. However, as shown in the gray region of Fig.~\ref{fig:TRRR33Jcp}, both phases are correlated and some part of the $\phi_1-\phi_2$ parameter space is excluded at $3\sigma$ of C.L for $1\,\text{d.o.f.}$. 

\begin{table}[t]
\begin{center}
\begin{tabular}{| c | c | c | }
\hline
Parameter & Best-fit & $3\sigma$ range \\ \hline \hline
$m_0$ ($\times 10^{-3}\text{eV}$) & $4.6$ & $[1.0, 6.1]$ \\ \hline
$a_n$ ($\times 10^{-2}\text{eV}$) & $3.0$ & $[0.5, 3.6]$ \\ \hline
$\phi_1/\pi$ & $0.2$ & $[-0.8, 0.8]$  \\ \hline
$\phi_2/\pi$ & $0.6$ & Unconstrained  \\ \hline 
\end{tabular}
\caption{\label{tab:rrr3fit} Best-fit parameters (second column) and three sigma allowed range for $1\,\text{d.o.f.}$ (third column) for each of the parameters of the $\text{RRR}_3$-form (first column) obtained from the minimization of Eq.~(\ref{eq:chi2NOexp}). The $\chi^2$ value at the minimum is $\chi^2_{\text{min.}}=0$.} 
\end{center}
\end{table}

\subsection{DUNE sensitivity to the mixing parameters}
\label{subsec:app2b}

\begin{figure}[t]
    \centering
    \includegraphics[width=\textwidth]{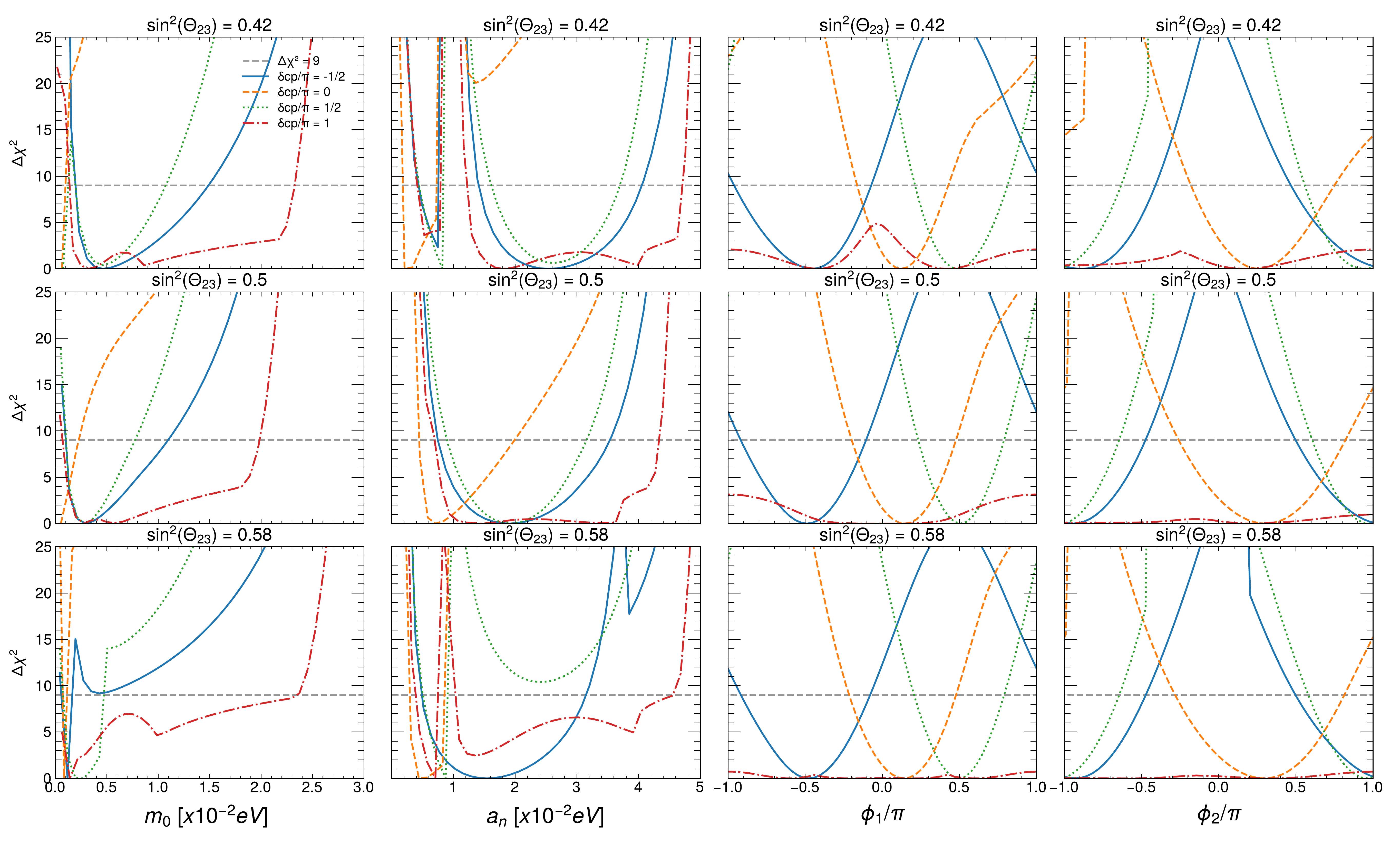}
    \caption{$\Delta \chi^2$ profiles for each one of the four mixing parameters $\vec{\lambda}=\{m_0, a_n, \phi_1, \phi_2\}$ that parametrize the mixing obtained from the $\text{RRR}_3$-form, after marginalization over the other three not shown parameters. The upper, middle, and lower row panels correspond to results for the input values $\sin^2{\theta}_{23}=\{0.42, 0.5, 0.58\}$, respectively. The lines correspond to results for the input values $\delta_{\text{CP}}/\pi=\{-0.5,0,0.5,1\}$, respectively. The horizontal dashed line corresponds to a $3\sigma$ of C.L for d.o.f. More details are in the text and in Table~\ref{tab:RRR33interval}. }
    \label{fig:RRR33layout}
\end{figure}

In this case we use simulated events at the DUNE FD to perform the analysis described in the main text (see Sec.~\ref{sec:expfit}). In Fig.\ref{fig:RRR33layout} we show the $\Delta \chi^2$ profiles for each one of the four $\vec{\lambda}$ parameters in terms of input values of the atmospheric mixing angle $\sin^2 \theta_{23}=\{0.42,0.5,0.58\}$ and $\delta_{\text{CP}}/\pi=\{-0.5,0,0.5,1\}$, while the remaining standard oscillation parameters were fixed to their best-fit values in Table~\ref{tab:oscparams}. In Table~\ref{tab:RRR33interval} we show the parameter ranges at $3\,\sigma$ of C.L for 1~d.o.f. from the sensitivity results in Fig.~\ref{fig:RRR33layout} for each one of the four $\vec{\lambda}$ parameters.

\begin{table}[t]
        \begin{tabular}{|c c| c c c c |c|}
        
            \hline
               $\sin^2(\theta_{23})$ &$ \delta_{\text{CP}}/\pi $& $ m_0\ (\times 10^{-2}~\text{eV})$ & $a_n$ $(\times 10^{-2}~\text{eV})$ & $\phi_1/ \pi$ &  $\phi_2 /\pi$ & $\chi^2_{\text{min.}}$ \\ [0.5ex] 
               \hline \hline
                  0.42 &  -0.5 &   [0.2,  1.7] &    $[0.2,  1.0]\cup [1.7, 4.1]$ &  [-1.0,  -0.1] &  [0.5,  1.0], [-1.5,  -0.4] &12.8\\
                  0.42 &   0.5 &   [0.1,  0.2] &  $[0.1,  0.4]$ &  [0.2,  0.8] &    [-1.0,  -0.7], [0.6,  1.0] &24.1\\
                  0.42 &     1 &   [0.2,  1.0] &  $[0.2,  1.0]\cup [1.9, 3.8]$ &   [-1.0,  1.0] &                  [-1.9,  1.0]   & 0\\
                  0.42 &     0 &   [0.2,  2.5] &  $[0.2,  1.0]\cup [1.3, 4.8]$ &   [-0.1,  0.4] &                  [-0.2,  0.7]   & 68.0\\
               \hline                              
                   0.5 &  -0.5 &   [0.1,  1.2]  &  [1.0,  3.5] &  [-0.9,  -0.1] &  $[0.5,  1.0], [-1.0,  -0.5]$ & 9.2\\
                   0.5 &     0 &    [0.0,  0.3]  &  [0.5,  2.0] &   [-0.1,  0.5] &                  [-0.2,  0.8] & 53.4 \\
                   0.5 &   0.5 &    [0.1,  0.9]  &  [1.1,  3.1] &    [0.2,  0.7] &  $[-1.0,  -0.7], [0.6,  1.4]$ & 17.3 \\
                   0.5 &     1 &   [0.1,  2.0]  &  [1.0,  4.4] &   [-1.0,  1.0] &                  [-1.0,  1.9] & 0\\
               \hline                              
                  0.58 &  -0.5 &    [0.1,  0.2]                &  [0.5,  3.1] &  [-0.9,  -0.1] & $ [-1.0,  -0.5], [0.5,  1.5]$ & 7.7\\
                  0.58 &     0 &    [0.1,  0.2]                &  [0.4,  0.7] &   [-0.2,  0.4] &                  [-0.2,  0.8] & 45.7\\
                  0.58 &   0.5 &    [0.1,  0.5]                &  [0.5,  1.0] &    [0.2,  0.8] &  $[0.6,  1.0], [-1.0,  -0.7 ]$ & 14.1 \\ 
                  0.5 &     1 &     $[0.0,  2.4]  \cup [1.7, 2.0]$ &  $[0.4,  0.7]\cup [1.2, 4.5]$ &   [-1.0,  1.0] &                  [-1.0,  1.0] & 0\\[1ex]
               \hline              
                        
        \end{tabular}              
        \caption{\label{tab:RRR33interval} Parameter ranges at $3\,\sigma$ of C.L for 1 d.o.f. from the sensitivity results in Fig.~\ref{fig:RRR33layout} for each one of the four mixing parameters $\vec{\lambda}=\{m_0, a_n, \phi_1, \phi_2\}$ that parametrize the mixing obtained from the $\text{RRR}_3$-form, for the different pair of input values of $\sin^2{\theta}_{23}$ and $\delta_{\text{CP}}$, as quoted in the first two columns of the table. The $\chi^2$ values at the bfp $\chi^2_{\text{min}}$, are also included in the last column. See main text for more details.}
        \end{table}
\clearpage
\bibliographystyle{apsrev4-1}
\bibliography{references}  

\end{document}